\documentclass[11pt]{myArticle}       

\usepackage{amsmath,amssymb,amsfonts} 
\usepackage{graphicx}
\usepackage{epsfig}
\usepackage{color}                    
\usepackage{hyperref}                 
\usepackage{myFancyhdr}               
\usepackage{makeidx}                  
\usepackage{helvet}                   
\usepackage{setspace}                 
\usepackage{fancybox}                 
\usepackage{float}                    
\usepackage{xspace}                   
\usepackage{wasysym}                  
\usepackage[numbers,sort&compress]{natbib}
\usepackage{hypernat}
\usepackage{wrapfig}
\usepackage{subcaption}
\usepackage{units}
\usepackage{lineno}
\usepackage{listings}

\definecolor{mygray}{rgb}{0.3,0.32,0.35}
\definecolor{darkblue1}{rgb}{0,0,.2}
\definecolor{darkblue}{rgb}{0,0,.3}
\definecolor{darkred}{rgb}{0.5,0,0}
\pagecolor{white} 
\color{black}     
\hypersetup{breaklinks=true,
            colorlinks=true,
            linkcolor=darkblue1,
            menucolor=darkblue1,
            urlcolor=darkblue1,
            citecolor=darkblue1,
            pdftitle={A new correlation coefficient between categorical, ordinal and interval variables with Pearson characteristics},
            pdfauthor={Max Baak},
            pdfsubject={A new correlation coefficient between categorical, ordinal and interval variables with Pearson characteristics},
            pdfkeywords={},
            pdfproducer={Advanced Analytics \& Big Data, KPMG Advisory N.V.}
}
\bibstyle{plain}
\renewcommand{\headrulewidth}{0.5pt}    
\renewcommand{\footrulewidth}{0.5pt}    

\newcommand\allFontSize{\small}

\newenvironment{myquote}
               {\list{}{\leftmargin0cm}%
                \item\relax}
               {\endlist}

\usepackage{caption}
\renewcommand{\captionfont}{\allFontSize}

\newcommand\detailsSize{\allFontSize}
\newenvironment{details}%
{\begin{myquote}\vspace{-0.2cm}\detailsSize}{\end{myquote}\vspace{-0.2cm}}

\newfloat{codeexample}{H}{loc}
\floatname{codeexample}{\sf\allFontSize Code Example}

\newlength{\gfitterboxwidth}
\definecolor{DarkGray}{rgb}{0.4,0.42,0.45}
\definecolor{LightGray}{rgb}{0.97,0.98,0.98}

{\VerbatimEnvironment\normalsize
\setlength{\gfitterboxwidth}{\textwidth}\addtolength{\gfitterboxwidth}{-2.3\fboxsep}%
\begin{Sbox}\begin{minipage}[t]{\gfitterboxwidth}\begin{Verbatim}}%
{\end{Verbatim}\end{minipage}\end{Sbox}\vspace{1ex} \fcolorbox{DarkGray}{LightGray}{\TheSbox}}

\newfloat{option}{H}{loc}
\floatname{option}{\sf\allFontSize Option Table}

\usepackage{tabularx}
{\setlength{\gfitterboxwidth}{\textwidth}\addtolength{\gfitterboxwidth}{-2.3\fboxsep}%
\begin{Sbox}\begin{tabular*}{\gfitterboxwidth}{>{\rule[-0.5ex]{.0ex}{3.0ex}\tt\small}p{3cm}>{\tt\small}p{4.5cm}>{\small}p{5.7cm}}
&&\\[-0.7cm]
\rm\small Option  & \rm\small Values    & \rm\small Description\\[0.1cm]\hline
&&\\[-0.45cm]}%
{\\[-0.1cm]\end{tabular*}\end{Sbox}\vspace{1ex} \fbox{\TheSbox}}

{\setlength{\gfitterboxwidth}{\textwidth}\addtolength{\gfitterboxwidth}{-2.3\fboxsep}%
\begin{Sbox}\begin{tabular*}{\gfitterboxwidth}{>{\rule[-0.5ex]{.0ex}{3.0ex}\tt\small}p{3cm}>{\tt\small}p{3.5cm}>{\small}p{6.7cm}}
&&\\[-0.7cm]
\rm\small Option  & \rm\small Values    & \rm\small Description\\[0.1cm]\hline
&&\\[-0.45cm]}%
{\\[-0.1cm]\end{tabular*}\end{Sbox}\vspace{1ex} \fbox{\TheSbox}}

{\setlength{\gfitterboxwidth}{\textwidth}\addtolength{\gfitterboxwidth}{-2.3\fboxsep}%
\begin{Sbox}\begin{tabular*}{\gfitterboxwidth}{>{\rule[-0.5ex]{.0ex}{3.5ex}\tt\small}p{3.0cm}>{\tt\small}p{3.0cm}>{\small}p{7.2cm}}
&&\\[-0.7cm]
\rm\small Option  & \rm\small Values    & \rm\small Description\\[0.1cm]\hline
&&\\[-0.45cm]}%
{\\[-0.1cm]\end{tabular*}\end{Sbox}\vspace{1ex} \fbox{\TheSbox}}

{\setlength{\gfitterboxwidth}{\textwidth}\addtolength{\gfitterboxwidth}{-2.3\fboxsep}%
\begin{Sbox}\begin{tabular*}{\gfitterboxwidth}{>{\rule[-0.5ex]{.0ex}{3.0ex}\tt\small}p{2.5cm}>{\tt\small}p{3.5cm}>{\small}p{7.2cm}}
&&\\[-0.7cm]
\rm\small Option  & \rm\small Values    & \rm\small Description\\[0.1cm]\hline
&&\\[-0.45cm]}%
{\\[-0.1cm]\end{tabular*}\end{Sbox}\vspace{1ex} \fbox{\TheSbox}}

{\setlength{\gfitterboxwidth}{\textwidth}\addtolength{\gfitterboxwidth}{-2\fboxsep}%
\begin{Sbox}\begin{tabular*}{\gfitterboxwidth}{>{\rule[-0.5ex]{.0ex}{3.0ex}\tt\small}p{4.5cm}>{\tt\small}p{3.2cm}>{\small}p{5.5cm}}
&&\\[-0.7cm]
\rm\small Option  & \rm\small Values    & \rm\small Description\\[0.1cm]\hline
&&\\[-0.45cm]}%
{\\[-0.1cm]\end{tabular*}\end{Sbox}\vspace{1ex} \fbox{\TheSbox}}

{\setlength{\gfitterboxwidth}{\textwidth}\addtolength{\gfitterboxwidth}{-2\fboxsep}%
\begin{Sbox}\begin{tabular*}{\gfitterboxwidth}{>{\rule[-0.5ex]{.0ex}{3.0ex}\tt\small}p{4.0cm}>{\tt\small}p{3.0cm}>{\small}p{6.2cm}}
&&\\[-0.7cm]
\rm\small Option  & \rm\small Values    & \rm\small Description\\[0.1cm]\hline
&&\\[-0.45cm]}%
{\\[-0.1cm]\end{tabular*}\end{Sbox}\vspace{1ex} \fbox{\TheSbox}}

\newfloat{programs}{H}{loc}
\floatname{programs}{\sf Table}

\usepackage{tabularx}
{\setlength{\gfitterboxwidth}{\textwidth}\addtolength{\gfitterboxwidth}{-2\fboxsep}%
\begin{Sbox}\begin{tabular*}{\gfitterboxwidth}{>{\rule[-0.5ex]{.0ex}{3.0ex}\tt\small}p{4.1cm}>{\small}p{9.5cm}}
&\\[-0.7cm]
\rm\small Macro & \rm\small Description\\[0.1cm]\hline
&\\[-0.45cm]}%
{\\[-0.1cm]\end{tabular*}\end{Sbox}\vspace{1ex} \fbox{\TheSbox}}

\newcommand{\phik}  {\ensuremath{\phi_K}\xspace}
\mathchardef\Upsilon="7107
\def\Y#1S{\ensuremath{\Upsilon{(#1S)}}\xspace}

\newcommand{\Kbar    }{\kern 0.2em\overline{\kern -0.2em K}{}\xspace}

\newcommand{\Kz      }{\ensuremath{K^0}\xspace}
\newcommand{\Kzb     }{\ensuremath{\Kbar^0}\xspace}
\newcommand{\KzKzb   }{\ensuremath{\Kz \kern -0.16em \Kzb}\xspace}
\newcommand{\Kp      }{\ensuremath{K^+}\xspace}
\newcommand{\Km      }{\ensuremath{K^-}\xspace}

\newcommand{\KpKm    }{\ensuremath{\Kp \kern -0.16em \Km}\xspace}

\newcommand\Dbar    {\kern 0.18em\overline{\kern -0.18em D}{}\xspace}

\newcommand\Bbar    {\kern 0.18em\overline{\kern -0.18em B}{}\xspace}

\newcommand\Bz      {\ensuremath{B^0}\xspace}

\newcommand\Bzb     {\ensuremath{\Bbar^0}\xspace}
\newcommand\Bu      {\ensuremath{B^+}\xspace}
\newcommand\Bub     {\ensuremath{B^-}\xspace}

\newcommand\BpBm    {\ensuremath{\Bu {\kern -0.16em \Bub}}\xspace}
\newcommand\Bs      {\ensuremath{B^0_{s}}\xspace}
\newcommand\Bsb     {\ensuremath{\Bbar^0_{s}}\xspace}

\newcommand\BzBzb   {\ensuremath{\Bz {\kern -0.16em \Bzb}}\xspace}
\newcommand\BszBszb {\ensuremath{\Bs {\kern -0.16em \Bsb}}\xspace}

\newcommand{\tev}{\ensuremath{\mathrm{Te\kern -0.1em V}}\xspace}
\newcommand{\gev}{\ensuremath{\mathrm{Ge\kern -0.1em V}}\xspace}
\newcommand{\mev}{\ensuremath{\mathrm{Me\kern -0.1em V}}\xspace}
\newcommand{\kev}{\ensuremath{\mathrm{ke\kern -0.1em V}}\xspace}
\newcommand{\ev}{\ensuremath{\mathrm{e\kern -0.1em V}}\xspace}
\newcommand{\gevc}{\ensuremath{{\mathrm{Ge\kern -0.1em V\!/}c}}\xspace}
\newcommand{\mevc}{\ensuremath{{\mathrm{Me\kern -0.1em V\!/}c}}\xspace}
\newcommand{\gevcc}{\ensuremath{{\mathrm{Ge\kern -0.1em V\!/}c^2}}\xspace}
\newcommand{\mevcc}{\ensuremath{{\mathrm{Me\kern -0.1em V\!/}c^2}}\xspace}

\newcommand{\bei}{\begin{itemize}}
\newcommand{\eei}{\end{itemize}}
\newcommand{\beq}{\begin{equation}}
\newcommand{\eeq}{\end{equation}}
\newcommand{\beqn}{\begin{eqnarray}}
\newcommand{\eeqn}{\end{eqnarray}}
\newcommand{\beqns}{\begin{eqnarray*}}
\newcommand{\eeqns}{\end{eqnarray*}}
\newcommand{\bitm}{\begin{itemize}}
\newcommand{\eitm}{\end{itemize}}

\def\@citex[#1]#2{\if@filesw\immediate\write\@auxout{\string\citation{#2}}\fi
  \@tempcnta\z@\@tempcntb\m@ne\def\@citea{}\@cite{\@for\@citeb:=#2\do
    {\@ifundefined
       {b@\@citeb}{\@citeo\@tempcntb\m@ne\@citea
        \def\@citea{,\penalty\@m\ }{\bf ?}\@warning
       {Citation `\@citeb' on page \thepage \space undefined}}%
    {\setbox\z@\hbox{\global\@tempcntc0\csname b@\@citeb\endcsname\relax}%
     \ifnum\@tempcntc=\z@ \@citeo\@tempcntb\m@ne
       \@citea\def\@citea{,\penalty\@m}
       \hbox{\csname b@\@citeb\endcsname}%
     \else
      \advance\@tempcntb\@ne
      \ifnum\@tempcntb=\@tempcntc
      \else\advance\@tempcntb\m@ne\@citeo
      \@tempcnta\@tempcntc\@tempcntb\@tempcntc\fi\fi}}\@citeo}{#1}}

\def\@citeo{\ifnum\@tempcnta>\@tempcntb\else\@citea
  \def\@citea{,\penalty\@m}%
  \ifnum\@tempcnta=\@tempcntb\the\@tempcnta\else
   {\advance\@tempcnta\@ne\ifnum\@tempcnta=\@tempcntb \else
\def\@citea{--}\fi
    \advance\@tempcnta\m@ne\the\@tempcnta\@citea\the\@tempcntb}\fi\fi}

\xspace

\definecolor{mygreen}{rgb}{0,0.6,0}
\definecolor{mygray}{rgb}{0.5,0.5,0.5}
\definecolor{mymauve}{rgb}{0.58,0,0.82}

\makeindex
\makeatletter
\newcommand*{\bdiv}{%
  \nonscript\mskip-\medmuskip\mkern5mu%
  \mathbin{\operator@font div}\penalty900\mkern5mu%
  \nonscript\mskip-\medmuskip
}
\makeatother

\begin{document}
%
%
\pagenumbering{arabic}
{\small
\color{mygray}
\begin{flushright}
{\sf\em \today} \\
\def\UrlFont{\sf\em}
\url{https://phik.rtfd.io}
\end{flushright}
}
\def\UrlFont{\rm}

\vspace{1.3cm}

{\sf\LARGE\bfseries
A new correlation coefficient between categorical, ordinal and interval variables with Pearson characteristics}

\vspace{1.0cm}

{\Large
  M.~Baak$^{a}$, R.~Koopman$^{a}$, H.~Snoek$^{b}$, S.~Klous$^{a,c}$
}

\vspace{0.5cm}

{\normalsize
  $^{a}$Advanced Analytics \& Big Data, KPMG Advisory N.V., Amstelveen, The Netherlands \\
  $^{b}$Nikhef National Institute for Subatomic Physics / University of Amsterdam, Amsterdam, The Netherlands \\
  $^{c}$Informatics Institute, University of Amsterdam, Amsterdam, The Netherlands

\vspace{1.0cm}

\begin{details} {\sf\bfseries Abstract}

A prescription is presented for a new and practical correlation coefficient,  \phik, based on several refinements to Pearson's hypothesis test of independence of two variables.
The combined features of  \phik form an advantage over existing coefficients. First, it works consistently between categorical, ordinal and interval variables.
Second, it captures non-linear dependency. Third, it reverts to the Pearson correlation coefficient in case of a bi-variate normal input distribution.
These are useful features when studying the correlation between variables with mixed types.
Particular emphasis is paid to the proper evaluation of statistical significance of correlations and to the interpretation of variable relationships
in a contingency table, in particular in case of low statistics samples and significant dependencies.
Three practical applications are discussed.
The presented algorithms are easy to use and available through a public Python library.
\end{details}

\thispagestyle{empty}

\newpage
%
%

\tableofcontents
\newpage

\section{Introduction}
\label{sec:intro}

The calculation of correlation coefficients between paired data variables is a standard tool of analysis for every data analyst.
Pearson's correlation coefficient~\cite{pearson_1895} is a \textit{de facto} standard in most fields, but by construction only works for interval variables. 
While many coefficients of association exist, each with different strengths, we have not been able to identify a correlation coefficient\footnote{The convention adopted here 
  is that a correlation coefficient is bound, \textit{e.g.} in the range $[0,1]$ or $[-1,1]$, and that a coefficient of association is not.} with
Pearson-like characteristics and a sound statistical interpretation that works for interval, ordinal and categorical variable types alike.

This paper describes a novel correlation coefficient, \phik, with properties that -- taken together -- form an advantage over existing methods.
Broadly, it covers three related topics typically encountered in data analysis:
\begin{enumerate}
\item Calculation of the correlation coefficient, \phik, for each variable-pair of interest.

The correlation \phik follows a uniform treatment for interval, ordinal and categorical variables.
This is particularly useful in modern-day analysis when studying the dependencies between a set of variables with mixed types,
where some variables are categorical.
The values for levels of correlation are bound in the range $[0,1]$, with $0$ for no association and $+1$ for complete association.
By construction, the interpretation is similar to Pearson's correlation coefficient, and is equivalent in case of a bi-variate normal input distribution.
Unlike Pearson, which describes the average linear dependency between two variables, \phik also captures non-linear relations.
Finally, \phik is extendable to more than two variables.
\item Evaluation of the statistical significance of each correlation.

The correlation \phik is derived from Pearson's $\chi^2$ contingency test~\cite{barnard_1992},
\textit{i.e.} the hypothesis test of independence between two (or more) variables in a contingency table,
henceforth called factorization assumption.
In a contingency table each row is the category of one variable
and each column the category of a second variable. Each cell describes the number of records occurring in both categories at the same time.
The asymptotic approximation commonly advertised to evaluate the statistical significance of the hypothesis test,
\textit{e.g.} by statistics libraries such as \texttt{R}~\cite{r.chisq.test} and \texttt{scipy}~\cite{scipy.stats.chi2contingency},
makes particular assumptions on the number of degrees of freedom and the shape of the $\chi^2$ distribution.
This approach is unusable for sparse data samples, which may happen for two variables with a strong correlation and for low- to medium-statistics
data samples, leading to incorrect $p$-values. (Examples follow in Section~\ref{sec:significance}.)
Presented here is a robust and practical statistical prescription for the significance evaluation of the level of variable association,
based on an adjustment of the $\chi^2$ distribution when using the $G$-test statistic~\cite{sokal_rohlf_2012}.
%
\item Insights in the correlation of each variable-pair, by studying outliers and their significances.
  
To help interpret any relationship found, we provide a method for the detection of
significant excesses or deficits of records with respect to the expected values in a contingency table, so-called outliers,
using a statistically independent evaluation for expected frequency of records.
We evaluate the significance of each outlier frequency, putting particular emphasis on
the statistical uncertainty on the expected number of records and on the scenario of low statistics data samples.
\end{enumerate}

The methods presented in this work can be applied to many analysis problems.
Insights in variable dependencies serve as useful input to all forms of model building, be it classification or regression based,
such as the identification of customer groups, outlier detection for predictive maintenance or fraud analytics,
and decision making engines.
More general, they can be used to find correlations across (big) data sets, and correlations over time (in correlograms).
Three use-cases are discussed, the study of numbers of insurance claims, survey responses, and clustering compatibility.

This document is organized as follows. 
A brief overview of existing correlation coefficients is provided in Section~\ref{sec:overview}.
Section~\ref{sec:pearson} describes the contingency test, 
which serves as input for Section~\ref{sec:phik}, detailing the derivation of the correlation coefficient \phik.
The statistical significance evaluation of the contingency test is discussed in Section~\ref{sec:significance}.
In Section~\ref{sec:outliers} we zoom in on the interpretation of the dependency between a specific pair of variables,
where the significance evaluation of outlier frequencies in a contingency table is presented.
Three practical applications of this can be
found in Section~\ref{sec:applications}.
Section~\ref{sec:code} describes the implementation of the presented algorithms in publicly
available Python code, 
before concluding in Section~\ref{sec:conclusion}.

\section{Measures of variable association}
\label{sec:overview}

A correlation coefficient quantifies the level of mutual, statistical dependence between two variables. 
Multiple types of correlation coefficients exist in probability theory, each with its own definition and features. 
Some focus on linear relationships where others are sensitive to any dependency, some are robust 
against outliers, etc. 
Typically their values range from $-1$ to $+1$ or $0$ to $+1$, where $0$ means no statistical association,
$+1$ means the strongest possible association, and $-1$ means the strongest negative relation. 
In general, different correlation coefficients are used to describe dependencies between interval, ordinal, and categorical variables.

This section briefly discusses existing correlations coefficients and other measures of variable association. This is done separately for interval, ordinal, and categorical variables.
In addition, several related concepts used throughout this work are presented.

An \textbf{\textit{interval variable}}, sometimes called continuous or real-valued variable, has well-defined intervals between the values of the variable.
Examples are distance or temperature measurements. 
The Pearson correlation coefficient is a \textit{de facto} standard to quantify the level of association between two interval variables.
For a sample of size $N$ with variables $x$ and $y$, it is defined as the covariance of the two variables divided by the product of their standard deviations: 
\begin{equation} \label{eq:pearsonrho}
\rho = \frac{\sum_{i=1}^{N}(x_i - \bar{x})(y_i - \bar{y})}{\sqrt{\sum_{i=1}^{N} (x_i - \bar{x})^2 }\sqrt{\sum_{i=1}^{N} (y_i - \bar{y})^2 }}\,,
\end{equation}
where $\bar{x}$ and $\bar{y}$ are the sample means.
Notably, $\rho$ is symmetric in $x$ and $y$, and $\rho \in [-1,1]$.

Extending this to a set of input variables, Pearson's correlation matrix $C$,
containing the $\rho$ values of all variable pairs, is obtained from the covariance matrix $V$ as:
\begin{equation}
C_{ij} = \frac{V_{ij}}{\sqrt{V_{ii}V_{jj}}}\,,
\end{equation}
where $ij$ are the indices of a variable pair.

The Pearson correlation coefficient measures the strength and direction of the linear relationship between two interval variables;
a well-known limitation is therefore that non-linear dependencies are not (well) captured.
In addition, $\rho$ is known to be to sensitive to outlier records.
Pearson's correlation coefficient, like many statistics formulas, requires interval variables as input, which can be unbinned or binned.
It cannot be evaluated for categorical variables, and ordinal variables can only be used when ranked (see below).

A direct relationship exists between $\rho$ and a bi-variate normal distribution:
\begin{align}
\label{eq:bivar}
f_{\mathrm{b.n.}} (x,y\, |\, \bar{x}, \bar{y}, \sigma_{x},\sigma_{y}, \rho) &=& \\
&& \!\!\!\!\!\!\!\!\!\!\!\!\!\!\!\!\!\!\!\!\!\!\!\!\!\!\!\!\!\!\!\!\!\!\!\!\!\!\!\!\!\!\!\!\!\!\!\!\!\!\!\!  \frac{1}{2\pi\sigma_{x}\sigma_{y}\sqrt{1-\rho^2}} \exp \Bigg( -\frac{1}{2(1-\rho^2)}\bigg[\frac{(x-\bar{x})^2}{\sigma_{x}^2} + \frac{(y-\bar{y})^2}{\sigma_{y}^2} - \frac{2\rho(x-\bar{x})(y-\bar{y})}{\sigma_{x}\sigma_{y}}\bigg] \Bigg)\,, \nonumber
\end{align}
where $\sigma_{x}$ ($\sigma_{y}$) is the width of the probability distribution in $x$ ($y$), and the correlation parameter
$\rho$ signifies the linear tilt between $x$ and $y$. 
We use this relation in Section~\ref{sec:phik} to derive the correlation coefficient \phik.

Another measure is the global correlation coefficient~\cite{james_roos_1975}, which is a number between zero and one obtained from the covariance matrix $V$ 
that gives the highest possible correlation between variable $k$ and the linear combination of all other variables:
\begin{equation} \label{eq:globalcorr}
g_k = \sqrt{ 1 - \big[ V_{kk} * (V^{-1})_{kk} \big]^{-1} }\,.
\end{equation}

An \textbf{\textit{ordinal variable}} has two or more categories with a clear ordering of these categories. 
For example, take the variable ``level of education'' with six categories: no education, elementary school graduate, high school graduate, college and university graduate, PhD.
A rank correlation measures the statistical relationship between two variables that can be ordered. 
The rank of a variable is its index in the ordered sequence of values. 
For ordinal variables a numbering is assigned to the categories, \textit{e.g.} 0, 1, 2, 3. Note the equidistant spacing between the categorical values.

Examples of rank correlation coefficients are Spearman's $\rho$~\cite{Spearman:1904}, Kendall's $\tau$~\cite{kendall_1938}, Goodman-Krustall's $\gamma$~\cite{Goodman:1954,Goodman:1959,Goodman:1963,Goodman:1972},
and the polychoric correlation~\cite{drasgow_2006}.
The definition of Spearman's $\rho$ is simply Eqn.~\ref{eq:pearsonrho}, using the ranks of $x_i$ and $y_i$ as inputs,
essentially treating the ranks as interval variables.
This makes Spearman's $\rho$ very robust against outliers.
Noteworthy, Goodman-Krustall's $\gamma$ is dependent on the order of the two input variables, resulting in an asymmetric correlation matrix.

Although ranking is regular practice, the assumption of equidistant intervals -- often made implicitly -- can sometimes be difficult to justify.
%
Adding the category of ``MBA'' to the above example increases the distance between ``PhD'' and ``no education'',
where one could argue that this distance should be independent of the number of educational categories.

A \textbf{\textit{categorical variable}}, sometimes called a nominal or class variable, has two or more categories which have no intrinsic ordering. 
An example is the variable gender, with two categories: male and female.
Multiple measures of association exist that quantify the mutual dependence between two (or more) categorical variables,
including Pearson's $\chi^2$ contingency test~\cite{barnard_1992}, the $G$-test statistic~\cite{sokal_rohlf_2012}, mutual information~\cite{cover_thomas_2006}, Fisher's exact test~\cite{fisher_1922,fisher_1970}, Barnard's test~\cite{Bernard:1945,barnard_1947}.
For an overview see Ref.~\cite{agresti_1992}.
These measures determine how similar the joint distribution $p(x,y)$ is to the product of the factorized marginal distributions $p(x)p(y)$. 
Each measure of association consists of a sum of contributions, one from each cell of the contingency table, and therefore does not depend
on the intrinsic ordering of the cells.

Though typically limited to categorical variables, these test statistics can also be applied to interval and ordinal type variables.
However, their values are not bound in the range $[0,1]$, and can become large.
Moreover, their interpretation is often complex, as their values not only depend on the level of association, but also on the number of categories or intervals and the number of records.

Most comparable to this work is Cram\'er's $\phi$~\cite{cramer_harald_1999}, a correlation coefficient meant for two categorical variables, denoted as $\phi_C$, 
based on Pearson's $\chi^2$ test statistic, and with values between $0$ (no association) and $+1$ (complete association):
\begin{equation}
\label{eq:phic}
\phi_{C} =  \sqrt{\frac{\chi^2}{N\min(r-1,k-1)}}\,,
\end{equation}
where $r$ ($k$) is the number of rows (columns) in a contingency table. 
Notably, with a relatively small number of records, comparable with the number of cells,
statistical fluctuations can result in large values of $\phi_C$ without strong evidence of a meaningful correlation.
(An example of this follows in Fig.~\ref{fig:rho0}a.)

Cram\'er's $\phi$ can also be used for ordinal and binned interval variables.
Fig.~\ref{fig:phic} shows $\phi_C$ for a binned bi-variate normal input distribution with correlation parameter $\rho$.
Compared to Pearson's $\rho$, $\phi_C$ shows relatively low values for most values of $\rho$, and only shoots up to one for values of $\rho$ close to one.
Moreover, the value found for $\phi_C$  is dependent on the binning chosen per variable,
as also seen in the figure.
This effect make $\phi_C$ difficult to interpret, and essentially unsuitable for interval variables.

\begin{figure}[htp]
  \centering
  \begin{minipage}[b]{0.6\linewidth}
    \centering
    \includegraphics[width=\textwidth]{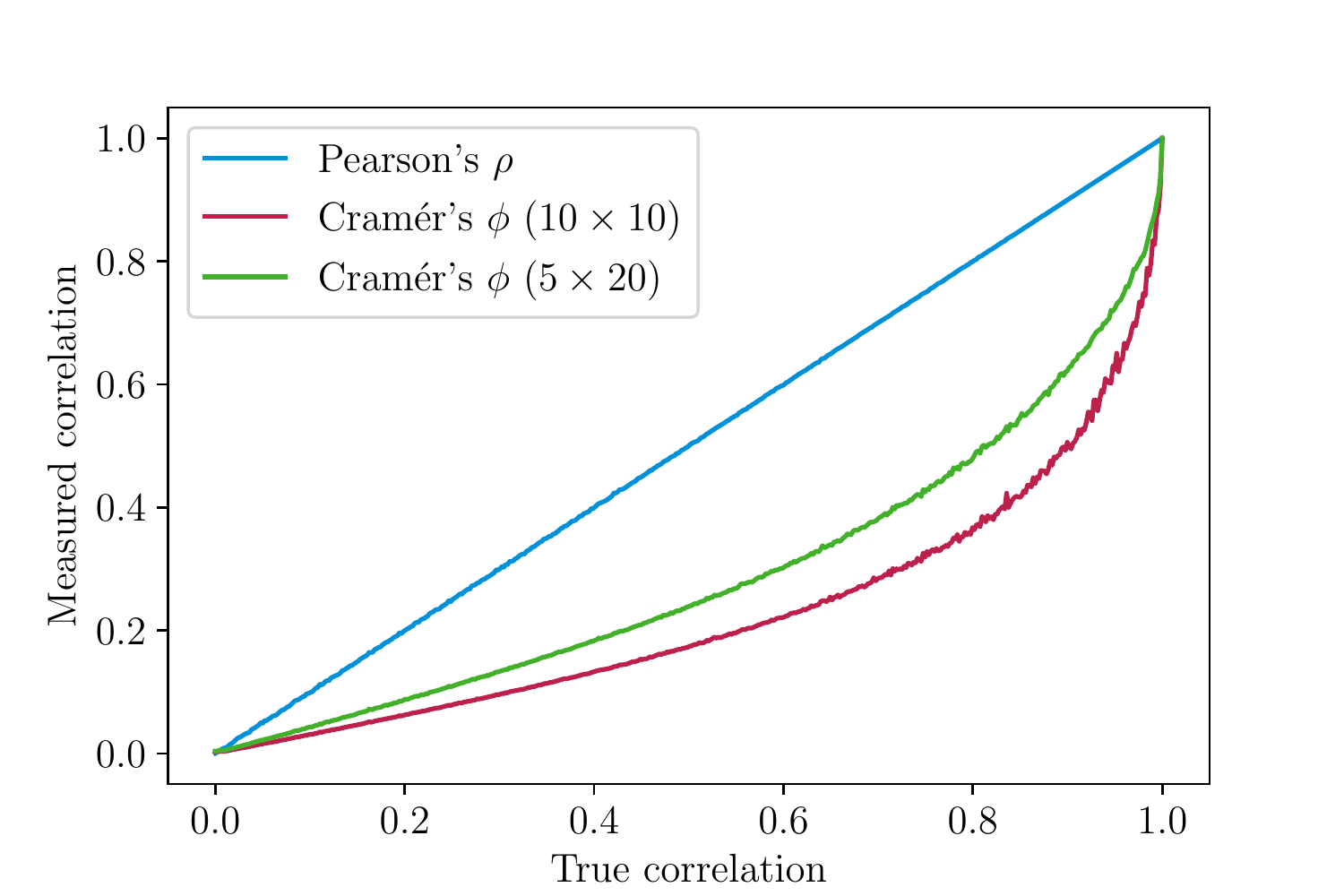}
  \end{minipage}%
  \caption{Cram\'er's $\phi$ versus Pearson's $\rho$.
    The two curves for Cram\'er's $\phi$ have been evaluated with different numbers of rows $r$ and columns $k$: $10\times10$ and $5\times20$ bins. The value found for $\phi_C$  is dependent on the number or rows and columns. Smaller values for $\phi_C$ are found for the case of equal number of rows and columns (red) compared with different number of rows and columns (green).}
  \label{fig:phic}
\end{figure}

One more alternative is the contingency coefficient $C_p$, which suffers from the disadvantage that its maximum value depends on the number of categories $r$ and $k$, and does not reach a maximum of one.
The recommendation~\cite{smith_2009} is not to use $C_p$ to compare correlations in tables with variables that have different numbers of categories
(\textit{i.e.} when $r \neq k$).

To address the aforementioned issues, in this paper we define the coefficient of correlation \phik,
derived from Pearson's $\chi^2$ contingency test in Section~\ref{sec:phik},
and its statistical significance, derived using the $G$-test in Section~\ref{sec:significance}.

\section{Test of variable independence}
\label{sec:pearson}

The contingency test, also called the test of variable independence,
determines if a significant relationship exists between two (or more) categorical variables. 
Though usually performed on two categorical variables, the test can equally be applied to ordinal and binned interval variables, 
and can be extended to an arbitrary number of variables.
Specifically, the contingency test indicates how well the joint data distribution $p(x,y)$ of variables $x$ and $y$ is described by the 
product of its factorized marginal distributions $p(x)p(y)$.

Throughout this paper we employ two contingency tests, where each
compares the observed frequency of each category for one variable with the expectation across the categories of the second variable:
\begin{enumerate}
\item Pearson's $\chi^2$ test: 
\begin{equation}
\label{eq:chi2}
\chi^2 = \sum_{i,j} \frac{(O_{ij} - E_{ij})^2}{E_{ij}}\,,
\end{equation}
which is used to define the correlation coefficient \phik in Section~\ref{sec:phik}.
Pearson's $\chi^2$ test is the standard test for variable independence.
\item The $G$-test, sometimes called log-likelihood ratio test:
\begin{equation}
\label{eq:gtest}
G = 2 \sum_{i,j} O_{ij} \log (O_{ij} / E_{ij}) \,,
\end{equation}
which is used to evaluate the significance of the contingency test in Section~\ref{sec:significance}. The sum is taken over all non-empty cells.
\end{enumerate}
In both formulas, $O_{ij}$ ($E_{ij}$) is the observed (expected) frequency of records for row $i$ and column $j$ of the contingency table.
The stronger the dependency between $x$ and $y$, the less well modeled is their distribution by the factorized distribution $p(x)p(y)$,
and the larger each test statistic value.

Under the factorization assumption, the expected frequencies can be obtained in two ways: statistically dependent and independent.

\subsection{Dependent frequency estimates}
\label{sec:depfreq}

The default method of frequency estimation for row $i$ and column $j$ includes $O_{ij}$, so $E_{ij}$ is statistically dependent on the observed 
frequency of its bin.

The expected value of the two nominal variables is calculated as:
\begin{equation}
\label{eq:dep_est}
E_{ij} = N\, p_{r}(i)\, p_{k}(j) = \frac{(\sum_{n=1}^{k}O_{in}) ( \sum_{m=1}^{r}O_{mj} )} {N}\,,
\end{equation}
where $p_{r}(i)$ ($p_{k}(j)$) is the $i^{\rm th}$ ($j^{\rm th}$) bin of the row-projected (column-projected) marginal probability mass function (p.m.f.) and $N$ is the number of records.
The statistical dependency between $E_{ij}$ and $O_{ij}$ arises as the expectation $E_{ij}$ for cell $ij$ includes the cell's observation $O_{ij}$ in both the sum over columns and rows, and as part of $N$.
The formula can be easily extended to an arbitrary number of variables.

We use Eqn.~\ref{eq:dep_est} for the definition of \phik in Section~\ref{sec:phik} and for the calculation of its significance in Section~\ref{sec:significance}, as this distribution matches the observed frequencies most closely.

\subsection{Independent frequency estimates}
\label{sec:indepfreq}

The second method of estimation of $E_{ij}$ excludes $O_{ij}$, \textit{i.e.} is statistically independent of the observed 
frequency of records for row $i$ and column $j$.
This estimate, known in high energy physics as the ABCD formula~\cite{Aaboud:2017nhr}, is given by:
\begin{equation}
\label{eq:abcd}
E_{ij} = \frac{B_{ij}\,C_{ij}}{D_{ij}} = \frac{(\sum_{n\neq j} O_{in}) ( \sum_{m\neq i} O_{mj} )} {  \sum_{m\neq i} \sum_{n\neq j}  O_{mn}  }\,,
\end{equation}
where by construction $O_{ij}$ is not part of $E_{ij}$, which allows for an objective comparison between observed and expected frequencies per bin.
This formula can also be extended to more variables, except that the denominator of Eqn.~\ref{eq:abcd}, which is different for each pair of indices, 
can easily become zero for low statistics samples.

Note that $B_{ij}$, $C_{ij}$, and $D_{ij}$ are sums of frequencies, each obeying Poisson statistics, and are statistically independent.
Consequently, the statistical uncertainty on $E_{ij}$ is evaluated with straight-forward error propagation~\cite{ku_1965} as:
\begin{equation}
\label{eq:abcderror}
\sigma_{E_{ij}}^{2} = \frac{\sigma_{B_{ij}}^2 C_{ij}^2}{D_{ij}^{2}} + \frac{\sigma_{C_{ij}}^2 B_{ij}^2}{D_{ij}^{2}}  + \frac{\sigma_{D_{ij}}^2 E_{ij}^2}{D_{ij}^{2}}\,.
\end{equation}
For an observed frequency of $Q$ records, $\sigma_{Q} = \sqrt{Q}$,
except when $Q=0$, in which case we set $\sigma_{Q} = 1$.
By doing so, when $B_{ij}$ or $C_{ij}$ is zero, and thus $E_{ij}=0$, this approach results in a non-zero error on $E_{ij}$.
The statistical uncertainty on the expected frequency, $\sigma_{E_{ij}}$, is only zero when both $B_{ij}$ and $C_{ij}$ are zero.

The expectation from Eqn.~\ref{eq:abcd} is built with fewer statistics than Eqn.~\ref{eq:dep_est} and thus is slightly less accurate.
Another difference is that the ABCD formula is not a true product of two (or more) factorized marginal distributions,
\textit{i.e.} the relative predictions for one row are not identical to those for another row, as is the case for dependent frequency estimates.

We use the independent frequency estimates of Eqn.~\ref{eq:abcd} for the detection of significant excesses or deficits of records over expected values in a contingency table in Section~\ref{sec:outliers}, for reasons described there.

\section{Definition of \phik}
\label{sec:phik}

The correlation coefficient \phik is obtained by inverting the $\chi^2$ contingency test statistic through the steps outlined below.
Although the procedure can be extended to more variables, the method is described with two variables for simplicity.

We define the bi-variate normal distribution of Eqn.~\ref{eq:bivar} with correlation parameter $\rho$ and unit widths, centered around the origin, and in the range $[-5,5]$ for both variables.
Using uniform binning for the two interval variables, with $r$ rows and $k$ columns, results in a corresponding bi-variate p.m.f..
With $N$ records, the observed frequencies, $O_{ij}$, are set equal to the probability per bin multiplied by $N$.
The expected frequencies $E_{ij}$, are set to the predictions from the bi-variate normal distribution with $\rho\!=\!0$, with $N$ records and the same binning.
We then evaluate the $\chi^2$ value of Eqn.~\ref{eq:chi2}.

Let us define this function explicitly.
First, we perform the integral of the bi-variate normal distribution over the area of bin $ij$
\begin{equation}
\label{eq:bnintegral}
F_{ij}(\rho) = \int_{\mathrm{area}_{\,ij}} f_{\mathrm{b.n.}} (x,y \,| \,\rho) \,{\rm d}x{\rm d}y \,,
\end{equation}
leading to the sum over bins:
\begin{equation}
\label{eq:chi2bn}
\chi^2_{\rm b.n.}(\rho,N,r,k) = N\, \sum_{i,\,j}^{k,\,r}  \frac{\left(F_{ij}\left(\rho=\rho\right) - F_{ij}\left(\rho=0\right)\right)^2} {F_{ij}\left(\rho=0\right)} \,.
\end{equation}
This $\chi^2$ value explicitly ignores statistical fluctuations in observed frequencies, and is a function of the numbers of rows and columns, $N$, and the value of $\rho$.

To account for statistical noise, we introduce a sample-specific pedestal related to
a simple estimate of the effective number of degrees of freedom of the bi-variate sample, $n_{\rm sdof}$:
\begin{equation}
\label{eq:simpleendof}
n_{\rm sdof} = (r-1)(k-1) - n_{\rm empty}({\rm expected})\,,
\end{equation}
with number of rows $r$ and columns $k$, and where $n_{\rm empty}({\rm expected})$ is the number of empty bins of the dependent frequency estimates of
the sample.
The pedestal is defined as:
\begin{equation} \label{eq:chi2min}
\chi^2_{\rm ped} = n_{\rm sdof} + c \cdot \sqrt{2 n_{\rm sdof}}\,.
\end{equation}
The noise pedestal is configurable through parameter $c$, and by default $c=0$.
See Section~\ref{sec:noise} for the impact of the noise pedestal on \phik and Section~\ref{sec:significance} for a discussion on the effective number of degrees of freedom.

The maximum possible $\chi^2$ value~\cite{cramer_harald_1999} of the contingency test is:
\begin{equation}
\label{eq:chi2_max}
\chi^2_{\rm max}(N,r,k) = N\min(r-1,k-1)\,,
\end{equation}
which depends only the number of records $N$, rows $r$, and columns $k$,
and is reached when there is a one-on-one dependency between the two variables.
Specifically note that $\chi^2_{\rm max}$ is independent of the shape of
distribution\footnote{Note that the $G$-test does not have this useful feature, making the $G$-test unsuitable for the calculation of \phik.} $p(x,y)$.
%

We scale Eqn.~\ref{eq:chi2bn} to ensure it equals $\chi^2_{\rm ped}$ for $\rho=0$ and $\chi^2_{\rm max}$ for $\rho=1$.
\begin{equation}
\label{eq:chi2bnscaled}
X^2_{\rm b.n.}(\rho,N,r,k) = \chi^2_{\rm ped} + \bigg\{ \frac{\chi^2_{\rm max}(N,r,k) - \chi^2_{\rm ped}}{\chi^2_{\rm b.n.}(1,N,r,k)} \bigg\} \cdot \chi^2_{\rm b.n.}(\rho,N,r,k) \,.
\end{equation}
This function is symmetric in $\rho$, and increases monotonically from $\chi^2_{\rm ped}$ to $\chi^2_{\rm max}$ as $\rho$ goes from zero to one.

We can now perform the necessary steps to obtain the correlation coefficient \phik:
\begin{enumerate}
\item In case of unbinned interval variables, apply a binning to each one. A reasonable binning is
  generally use-case specific. As a default setting we take $10$ uniform bins per variable.
\item Fill the contingency table for a chosen variable pair, which contains $N$ records, has $r$ rows and $k$ columns.
\item Evaluate the $\chi^2$ contingency test using the Pearson's $\chi^2$ test statistic (Eqn~\ref{eq:chi2}) and the statistically dependent frequency estimates, as detailed in Section~\ref{sec:depfreq}.
\item Interpret the $\chi^2$ value as coming from a bi-variate normal distribution without statistical fluctuations, using Eqn.~\ref{eq:chi2bnscaled}.
\begin{itemize}
\item If $\chi^2 < \chi^2_{\rm ped}$, set $\rho$ to zero. 
\item Else, with fixed $N$, $r$, $k$, invert the $X^2_{\rm b.n.}$ function, \textit{e.g.} using Brent's method~\cite{brent_1973}, and numerically solve for $\rho$ in the range $[0,1]$.
\item The solution for $\rho$ defines the correlation coefficient \phik.
\end{itemize}
\end{enumerate}

The procedure can be extended to more variables by using a multi-variate Gaussian instead of a bi-variate one.

In summary, we interpret the $\chi^2$ value found in data as coming from a bi-variate normal distribution
with a fixed amount of statistical noise
and with correlation parameter \phik.
Non-linear relations are captured by \phik through the $\chi^2$ test of variable independence.
The correlation \phik reverts to the Pearson correlation coefficient in case of a bi-variate normal input distribution, with uniformly binned interval variables.
Unlike Cram\'er's $\phi$, the value of \phik is stable against the number of bins chosen per interval variable,
making it unambiguous to interpret.
(In Fig.~\ref{fig:phic}, overlaying the \phik values evaluated with (a)symmetric binning gives a line indistinguishable from Pearson's $\rho$.)
Like Cram\'er's $\phi$, \phik is affected by statistical fluctuations, which is relevant when the number of records
is comparable with the number of cells (or lower); 
however, unlike Cram\'er's $\phi$, \phik has a correction for the statistical noise.
Note that \phik is independent of the order of the two input variables,
and that the procedure can be extended to more than two variables\footnote{For more than two variables, follow the same procedure and assume a common correlation for each variable pair of the multivariate normal input distribution.}.

All coefficients presented in Section~\ref{sec:overview} are computationally inexpensive to evaluate.
The calculation of \phik is computationally expensive because of the integrals of correlated bi-variate normal distributions
evaluated in Eqn.~\ref{eq:chi2bnscaled}, but is well-doable on any modern laptop, typically taking only a fraction of a second per \phik calculation.

\subsection{Performance on benchmark samples}

A comparison with alternative correlation coefficients based on benchmark samples is given in Fig.~\ref{fig:benchmark}.
By construction, the interpretation of \phik is similar to that of Pearson's correlation coefficient,
in particular for the bi-variate normal input distributions and the linear shapes, shown in the left and middle columns. 
Unlike Pearson, however, \phik also captures non-linear relations as shown in the right column. Moreover, \phik can be determined for categorical, ordinal, and interval variables alike.
Note that Cram\'er $\phi$ gives relatively low values for all samples.

\begin{figure}[htp]
  \centering
  \begin{minipage}[b]{0.75\linewidth}
    \centering
    \includegraphics[width=\textwidth]{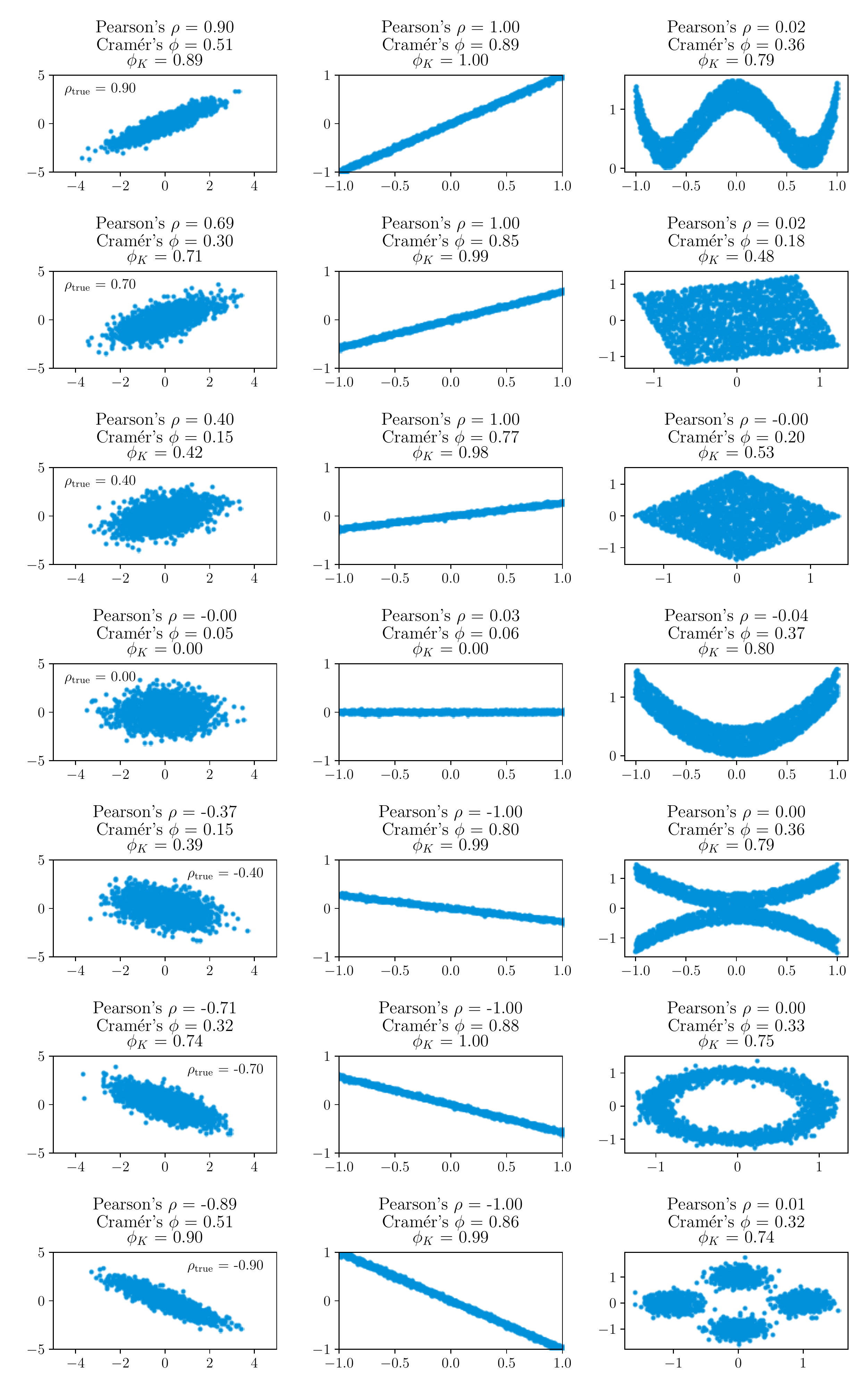}
  \end{minipage}%
  \caption{Benchmark sample results for \phik. Each synthetic data set contains 2000 data points. For the left column, from top to bottom, the bi-variate normal distributions have been generated
    with true correlations: $\{0.9,0.7,0.4,0,-0.4,-0.7,-0.9\}$. For the middle column a linear data set is generated which is rotated around the origin. In the right column various data sets are generated with non-linear correlations. Note that these non-linear correlations are well-captured by \phik, while Pearson's $\rho$ is close to zero for all cases.}
  \label{fig:benchmark}
\end{figure}

\subsection{Example correlation matrix}

When studying the dependencies of a set of variables with a mixture of types,
one can now calculate the correlation matrix for all variable pairs, filled with \phik values,
which is a useful overview to have for a data analyst.

For illustration purposes a synthetic data set with car insurance data has been created. 
The data set consists of 2000 records. 
Each record contains 5 (correlated) variables of mixed variable types, see Table~\ref{tab:data}.
These data are used throughout the paper to provide insights in the practical application of the methods introduced in this work. 
The \phik correlation matrix measured on the car insurance data set is shown in Fig.~\ref{fig:phik_example}. 

\begin{table}
\centering
\begin{minipage}[t]{0.55\textwidth}
\small
\begin{tabular}{rrrrr}
\hline
  car color & driver age &      area & mileage & car size \\
\hline
       blue &       60.4 &       suburbs &     3339 &       XS \\
       blue &       30.9 &       suburbs &    53370 &       XL \\
       blue &       18.5 &       suburbs &   112557 &       XL \\
      green &       40.9 &      downtown &    29605 &        L \\
       gray &       23.7 &      downtown &    15506 &        M \\
 multicolor &       60.3 &      downtown &    33148 &        L \\
      white &       66.7 &       suburbs &    91132 &       XL \\
        red &       69.2 &      downtown &   152445 &      XXL \\
   metallic &       43.5 &         hills &   147275 &        S \\
      ... & ... & ... & ... & ... \\
\hline
\end{tabular}
\end{minipage}
\caption{A synthetic data set with car insurance data. The data set consists of 2000 records and is used to illustrate the calculations of \phik, statistical significance (in Section~\ref{sec:significance}) and outlier significance (in Section~\ref{sec:outliers}).}
\label{tab:data}
\end{table}

\begin{figure}[htp]
\captionsetup[subfigure]{position=b}
\centering
    \begin{subfigure}[t]{0.55\textwidth}
       \vspace{0pt}
       \includegraphics[width=\linewidth]{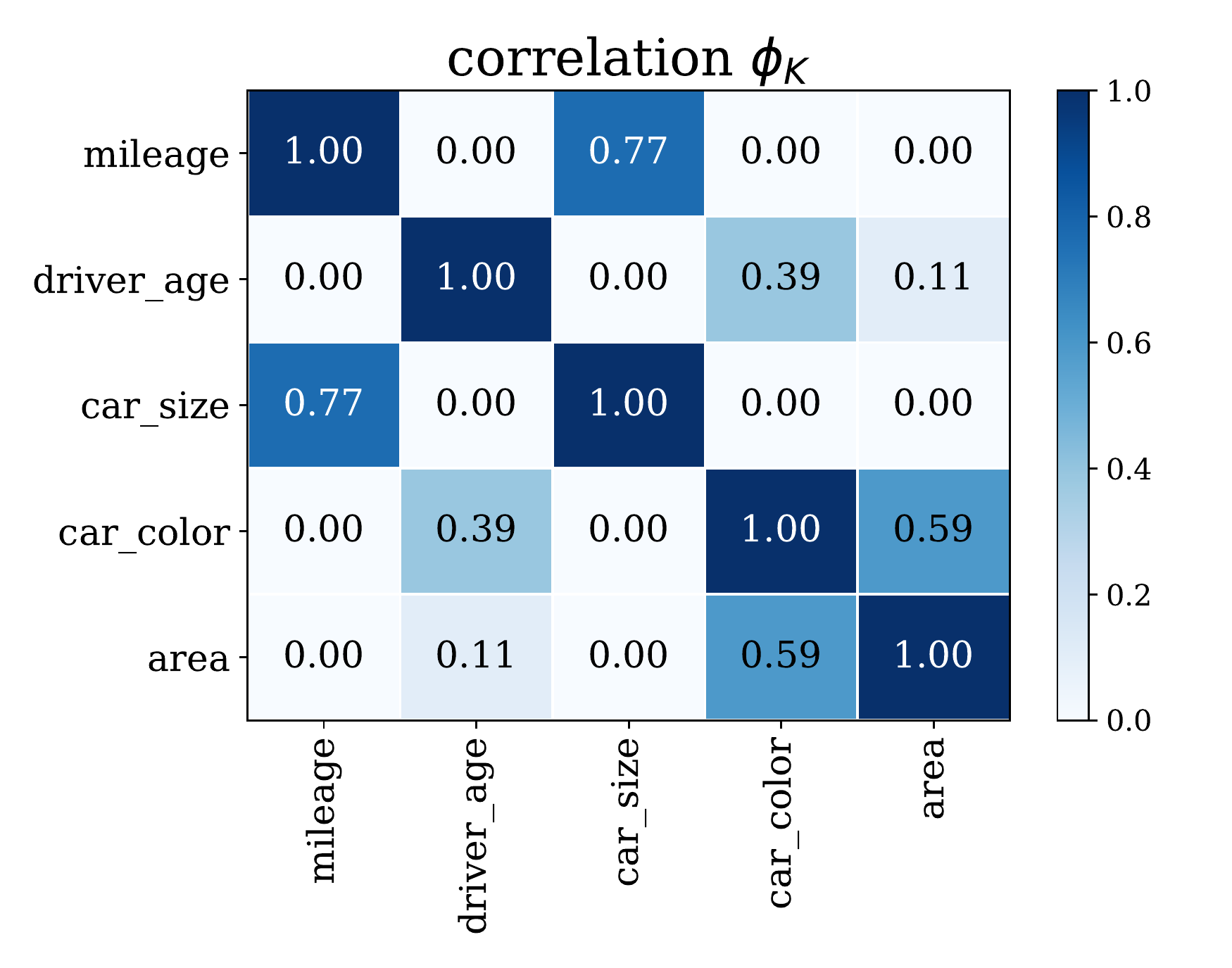}
       \caption{ }
    \end{subfigure}
    \hspace{0pt}
    \begin{subfigure}[t]{0.32\textwidth}
    \centering
       \vspace{0pt}
       \includegraphics[width=\linewidth]{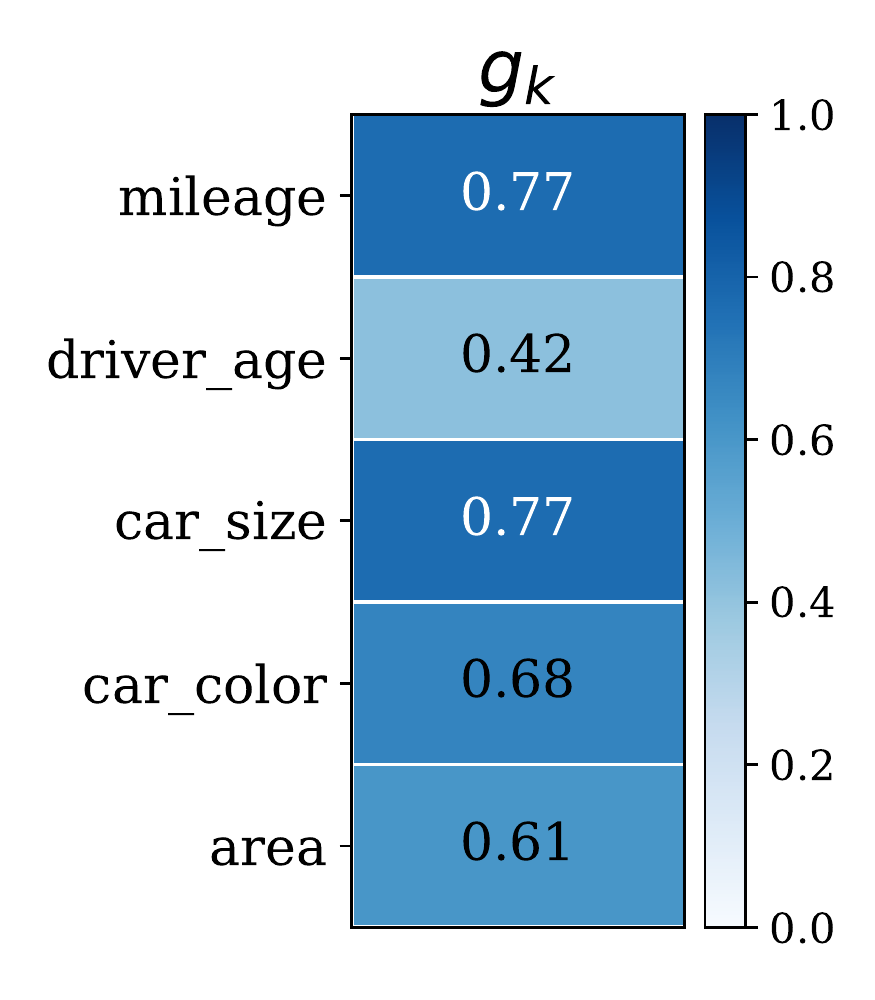}
       \vspace{18pt}
       \caption{}
     \end{subfigure}
      \caption{Correlation coefficients calculated on the synthetic car insurance data set (Table~\ref{tab:data}) containing mixed variables types. a) The \phik correlation matrix. b) The global correlations $g_k$.}
  \label{fig:phik_example}
\end{figure}

\subsection{Global correlation coefficients}

Besides the variable-pair information available from the correlation matrix $C$ in Fig.~\ref{fig:phik_example}a, it is also interesting to evaluate per variable the
global correlation coefficient, $g_k$, of Eqn.~\ref{eq:globalcorr}.
Strictly speaking, $g_k$ is only defined for interval variables, as it requires a covariance matrix $V$.
Here, we set the variances of all variable types to one (anyhow undefined for categorical variables\footnote{Interval
  variable can always be re-scaled to have unit variance.}) and use $V = C$.
Example global correlations measured in the car insurance data are shown in Fig.~\ref{fig:phik_example}b.
They give a tenable estimate of how well each variable
can be modeled in terms of all other variables, irrespective of variable type.

\subsection{Statistical noise correction} \label{sec:noise}

The calculation of \phik contains a correction for statistical fluctuations:
for any $\chi^2$ value below the sample-specific noise threshold $\chi^2_{\rm ped}$ of Eqn.~\ref{eq:chi2min},
indicating that no meaningful correlation can be determined, \phik is set to $0$ by construction.

\begin{figure}[htp]
  \centering
  \begin{subfigure}[t]{0.5\linewidth}
    \centering
    \includegraphics[width=\textwidth]{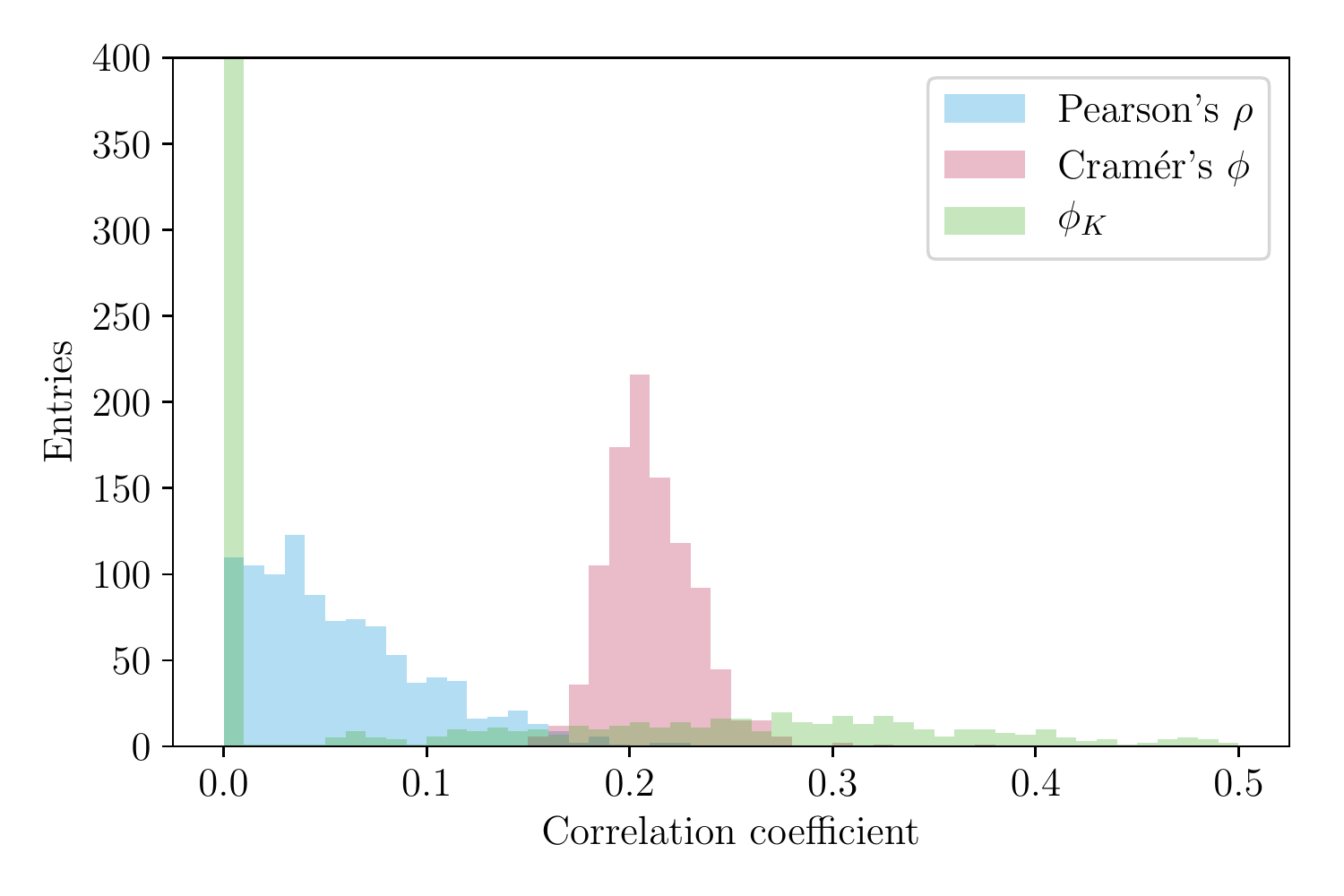}
    \caption{}
  \end{subfigure}%
  \begin{subfigure}[t]{0.5\linewidth}
    \centering
    \includegraphics[width=\textwidth]{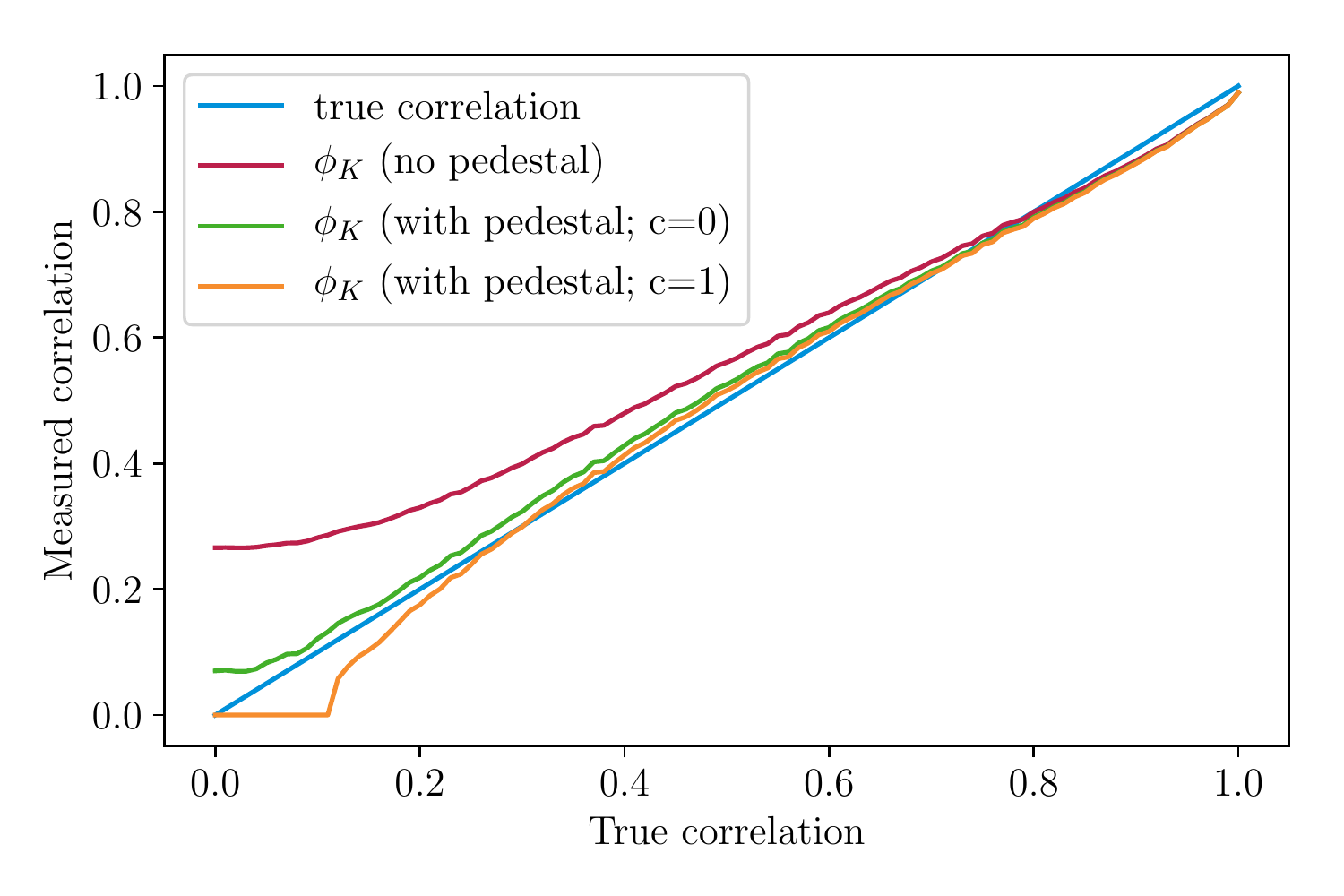}
    \caption{}
  \end{subfigure}
  \caption{a) The correlation coefficients Pearson's $\rho$, Cram\'er's $\phi$ and \phik, measured for 1000 synthetic data sets with 500 data points each, which are simulated using a bi-variate normal with parameter $\rho=0$. The absolute value of the Pearson's $\rho$ is taken as the measured $\rho$ can also take on negative values. 
    b) The median \phik value measured in 1000 synthetic data sets containing 500 data points each, simulated using a bi-variate normal distribution, as a function of true correlation $\rho$. The value of \phik is evaluated using three different configurations of the noise pedestal parameter $c$ (see Eqn.~\ref{eq:chi2min}).}
  \label{fig:rho0}
\end{figure}

The impact of the noise correction is seen in Fig.~\ref{fig:rho0}a, showing the absolute value of Pearson's $\rho$, Cram\'er's $\phi$, and \phik 
measured for 1000 synthetic data sets with only 500 records each,
simulated from a bi-variate normal distribution with no correlation, and each binned in a 10x10 grid.
Without absolute function applied, the distribution of Pearson's $\rho$ values would be centered around zero, as expected.
The calculation of Cram\'er's $\phi$ results in a seemingly significant bump at $0.2$.
This cannot be interpreted as a meaningful correlation, but results from the statistical noise contributing to each sample's $\chi^2$ value.

For \phik, only when $\chi^2 > \chi^2_{\rm ped}$ does the calculation of \phik kick into action.
The noise threshold is set such that about 50\% of the simulated samples gets assigned $\phik=0$.
The remaining samples result in a wide distribution of \phik values\footnote{Without noise correction,
  the \phik distribution shows a similar peak as Cram\'er's $\phi$, at value 0.5.}.

Fig.~\ref{fig:rho0}b shows \phik as a function of true correlation,
where \phik is obtained from the median $\chi^2$ value of 1000 synthetic data sets with 500 data points each.
The median gives the most representative, single synthetic data sample.
In the calculation of \phik three configurations for the noise pedestal of Eqn.~\ref{eq:chi2min} are tested:
no pedestal, and $c\in\{0,1\}$.
No pedestal gives \phik values that overshoot the true correlation significantly
at low values.
Configuration $c=1$ undershoots: the calculation of \phik turns on too late.
Configuration $c=0$ follows the true correlation most closely.
The residual differences disappear for larger data samples, and we deem this acceptable for this level of statistics
(with on average only 5 records per bin).

The sample-specific noise threshold $\chi^2_{\rm ped}$ depends mostly on the number of filled cells, and stabilizes
for larger sample sizes.
Consequently, its impact is rather limited for large samples with a meaningful, non-zero correlation,
typically having $\chi^2 \gg \chi^2_{\rm ped}$.
For small sample sizes, as is also obvious from Fig.~\ref{fig:rho0}b, any correlation coefficient value should first be 
held up against the significance of the hypothesis test of variable independence -- the topic of Section~\ref{sec:significance} -- before further interpretation.

\section{Statistical significance}
\label{sec:significance}

In practice, when exploring a data set for variable dependencies, the studies of correlations and their significances are equally
relevant: a large correlation may be statistically insignificant, and vice versa a small correlation may be very significant.

Both Pearson's $\chi^2$ test and the $G$-test asymptotically approach the $\chi^2$ distribution~\cite{barnard_1992}. 
For samples of a reasonable size (Cochran's rule on what defines ``reasonable size'' follows below), the default approach to obtain the $p$-value for the hypothesis test of variable independence is to 
integrate the $\chi^2$ probability density function $g(x|k)$ over all values\footnote{The integral runs up to infinity,
  even though the contingency test has a maximum test statistic value. In practice the difference is negligible.}
equal to or greater than the observed test statistic value $t_{\rm obs}$:
\begin{equation} \label{eq:pvalue}
p = \int_{t_{\rm obs}}^{\infty} g(x| k) \,{\rm d}x\,,
\end{equation}
with the p.d.f. of the $\chi^2$ distribution:
\begin{equation} \label{eq:chi2dist}
g(x | k) = \frac{1}{2^{\mu}\Gamma(\mu)}\cdot x^{\mu-1} \cdot e^{-x/2} \,,
\end{equation}
where $\mu = k/2$, $\Gamma(\mu)$ is the gamma function, and $k$ is set to the number of degrees of freedom $n_{\rm dof}$.
The solution of this integral is expressed as the regularized gamma function. This approach holds for samples of a reasonable size, and when using the $\chi^2$ test statistic or $G$-test.


For the independence test of $n_{\rm dim}$ variables, the number of degrees of freedom
is normally presented~\cite{bock_velleman_d._2007} as
the difference between the number of bins $n_{\rm bins}$ and model parameters $n_{\rm pars}$
\begin{eqnarray}
\label{eq:ndof}
n_{\rm dof} &=& n_{\rm bins} - n_{\rm pars} \nonumber \\
           &=& \Bigg[ \prod_{i=1}^{n_{\rm dim}} n_i \Bigg]  - \Bigg[ \sum_{i=1}^{n_{\rm dim}} (n_i \!-\! 1)\ +\ 1 \Bigg] \,.
\end{eqnarray}
where $n_i$ is the number of categories of variable $i$.
Explained using Eqn.~\ref{eq:dep_est}, each dimension requires $(n_i - 1)$ parameters to model its p.m.f.,
which is normalized to one, and the p.m.f. product is scaled to the total number of events, which requires one more parameter.
For just two variables this reduces to:
\begin{equation}
\label{eq:ndof2d}
n_{\rm dof} = (r-1)(k-1)\,.
\end{equation}

In practice Eqn.~\ref{eq:ndof} does not hold for many data sets, in particular for distributions with unevenly filled or unfilled bins.
For example, in the case of two (binned) interval variables with a strong dependency.
The \textit{effective} number of degrees of freedom, $n_{\rm edof}$, is often smaller than the advocated value, $n_{dof}$, and can even take on floating point values,
because the number of available bins is effectively reduced.


The asymptotic approximation, Eqns.~\ref{eq:pvalue}-\ref{eq:chi2dist}, 
breaks down for sparse data sets,
for example for two (interval) variables with a strong correlation, and for low-statistics data sets.
The literature on evaluating the quality of this approximation is extensive; for an overview see Refs.~\cite{agresti_2001,kroonenberg_verbeek_2018}.
Cochran's rule of thumb is that at least $80\%$ of the expected cell frequencies is $5$ counts or more, and that no expected cell frequency is less than $1$ count.
For a 2x2 contingency table, Cochran recommends~\cite{Cochran:1952,cochran_1954} that the test should be used only if the expected frequency in
each cell is at least $5$ counts.


How to properly evaluate the $p$-value 
if the test statistic does not follow the $\chi^2$ distribution and hence Eqn.~\ref{eq:pvalue} cannot be be safely applied. 
A reasonable approach is to evaluate Eqn.~\ref{eq:pvalue} directly with Monte Carlo data sets, sampled randomly from the distribution of expected frequencies.
However, this approach quickly becomes cumbersome for $p$-values smaller than $0.1\%$, \textit{i.e.} once more than 1000 simulations are needed for a decent $p$-value estimate,
and practically impossible when needing at least a million simulations.
Given that variable dependencies can be very significant,
we prefer a common approach that works for both strong and weak dependencies and both low- and high-statistics samples.

In this section we propose another option: a hybrid approach where a limited number of Monte Carlo simulations is used to fit
an analytical, empirical description of the $\chi^2$ distribution. 
Specifically, we describe two corrections to Eqn.~\ref{eq:pvalue}: 
\begin{enumerate}
\item A procedure to evaluate the effective number of degrees of freedom for a contingency test;
\item A correction to Eqn.~\ref{eq:chi2dist} for low statistics samples, when using the $G$-test statistic.
\end{enumerate}
We conclude the section with a prescription to evaluate the statistical significance of the hypothesis test of variable independence,
and a brief overview of sampling methods to help evaluate the $p$-value.

\subsection{Effective number of degrees of freedom} \label{sec:nedof}

To obtain the effective number of degrees of freedom of any sample, we use the property of Eqn.~\ref{eq:chi2dist} that,
for a test statistic distribution obeying $g(x|k)$,
to good approximation the average value of $g(x|k)$ equals $k$.

The effective number of degrees of freedom for any sample is obtained as follows:
\begin{enumerate}
\item For the two variables under study, the dependent frequency estimates form the factorized distribution most accurately describing the observed data.
  Using Monte Carlo sampling techniques, this distribution is used to randomly generate $500$ independent synthetic data sets with the same number of records
  as in the observed data set.

  Optionally, sampling with fixed row and/or column totals may be chosen.
  A short discussion of sampling methods is held in Section~\ref{sec:simapp}.
\item For each synthetic data set, evaluate the $G$-test statistic using the statistically dependent frequency estimates,
  as detailed in Section~\ref{sec:pearson}.
\item The effective number of degrees of freedom, $n_{\rm edof}$, is taken as the average value of the $G$-test distribution of all generated Monte Carlo samples.
\end{enumerate}

\begin{figure}[htp]
  \centering
  \begin{minipage}[b]{0.7\linewidth}
    \centering
    \includegraphics[width=\textwidth]{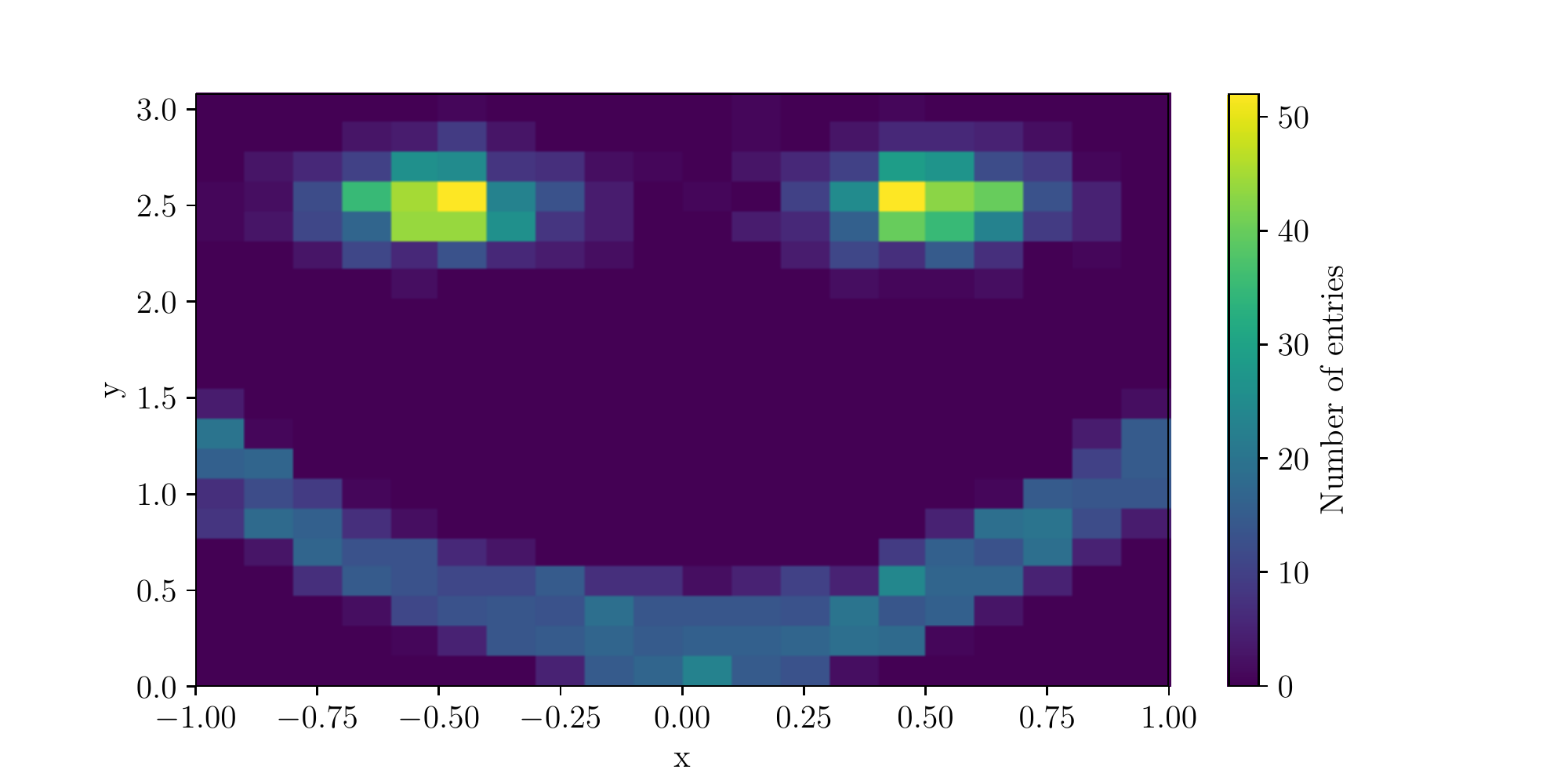}
  \end{minipage}%
  \caption{Example ``smiley'' data set of two interval variables binned in 20 bins in the $x$ and $y$ direction.}
  \label{fig:smiley}
\end{figure}

\begin{figure}[htp]
  \centering
  \begin{minipage}[b]{0.7\linewidth}
    \centering
    \includegraphics[width=\textwidth]{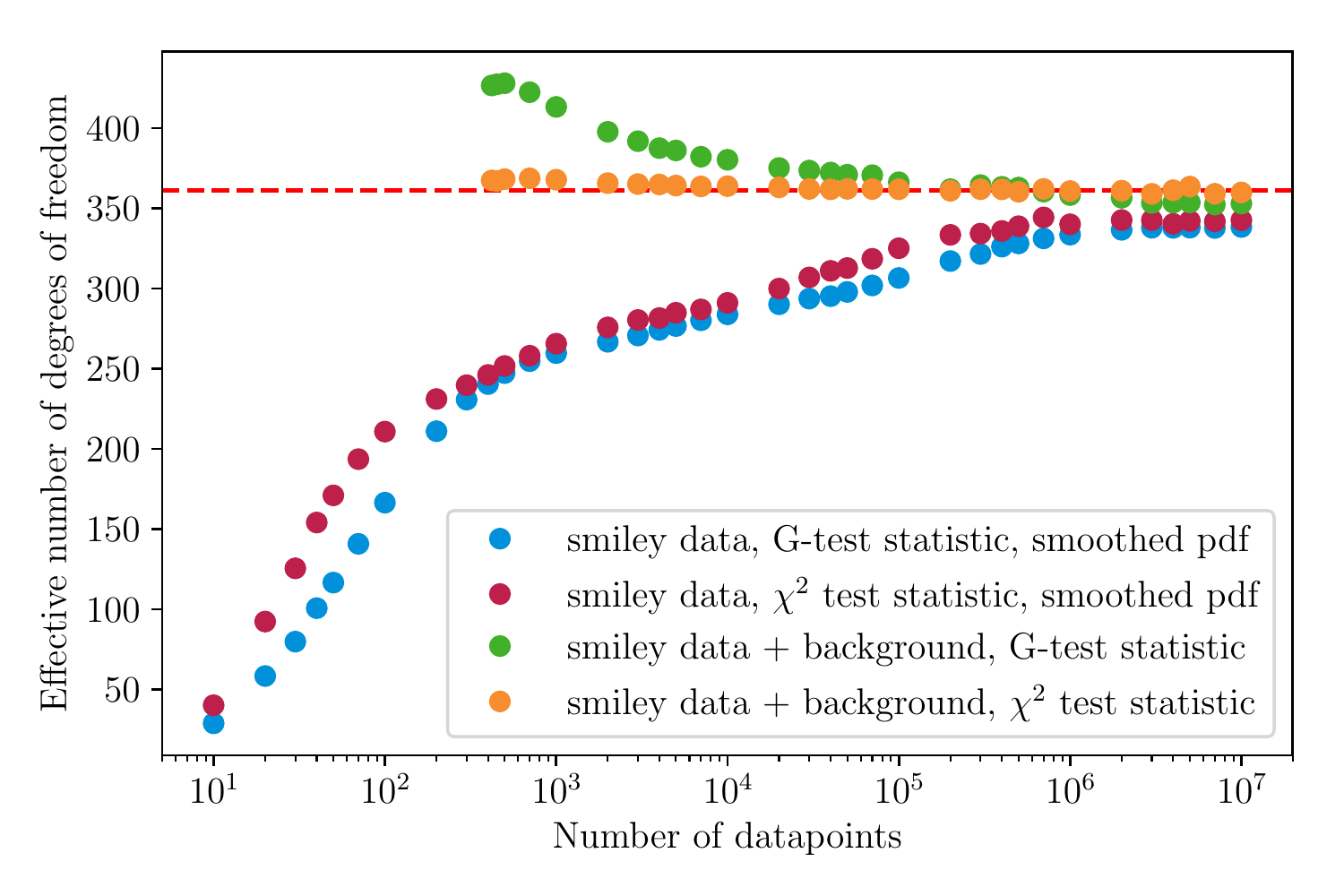}
  \end{minipage}%
  \caption{The effective number of degrees of freedom as a function of the number of data points in the input data set (Figure~\ref{fig:smiley}). The theoretical number of degrees of freedom, $n_{dof} = 361$, is indicated with the dashed line.}
  \label{fig:nedof}
\end{figure}

Fig.~\ref{fig:smiley} shows a ``smiley'' data set of two interval variables, consisting of two blobs and a wide parabola, which are binned into a 20x20 histogram,
for which we can generate an arbitrary number of records.

The bottom two curves in Fig.~\ref{fig:nedof} show $n_{\rm edof}$ obtained for this sample,
as a function of the number of records in the data set, $N$, and evaluated using the $G$-test and $\chi^2$ test statistic.
Using Eqn.~\ref{eq:ndof2d}, the advocated number of degrees of freedom of this sample equals $361$.
For both test statistics this number is only reached for very large sample sizes ($\ge 10^6$).
and drops significantly for smaller values of $N$, where the drop is slightly steeper for the $G$-test statistic.
The top two curves show the same data set on top of a uniform background of 1 record per cell, ensuring that each is always filled,
again evaluated using the $G$-test or $\chi^2$ test statistic.
Now the $G$-test overshoots, and the $\chi^2$ test statistic happens to level out at the expected value.

To understand the behavior of under- and overshooting, realize that $n_{\rm edof}$ relates directly to the distribution of dependent frequency estimates.
By construction, the dependent frequency estimates of Eqn.~\ref{eq:dep_est} make non-zero predictions for each bin in the distribution,
as long as the input data set contains at least one record per row and column.
Under the assumption of variable independence, each bin in the distribution is expected to be filled.

First consider the bottom two curves of Fig.~\ref{fig:nedof}.
For an uneven input distribution, for example two strongly correlated interval variables,
one may expect many bins with low frequency estimates.
A data set sampled randomly from a distribution with very low frequency estimates, such as the data set in Fig.~\ref{fig:smiley}, is likely to contain empty bins.
On average, high-statistics bins contribute $n_{\rm dof} / n_{\rm bins}\ (\lesssim 1)$ to the $G$-test or $\chi^2$ test statistic,
but the low-statistics bins do not obey this regime.
As an example, let us focus on the empty bins.
By construction their contribution to the $G$-test is zero.
The contribution to the $\chi^2$ test statistic is non-zero: $\sum_i E_i$, where the sum runs over all empty bins.
It is clear however, when $E_{i} \ll 1$, that this sum is relatively small and contributes only marginally.
Taken over many randomly sampled data sets, this effect reduces the average value of the $G$-test or $\chi^2$ test statistic distribution
to lower values, and likewise decreases $n_{\rm edof}$ compared with $n_{\rm dof}$.

For the top two curves, by construction $E_{i} > 1$ for each bin, bringing them closer to the nominal regime
and increasing the $G$-test and $\chi^2$ test statistics.
For a discussion of the contribution of low-statistics contingency table cells to the $\chi^2$ test statistic,
see Ref.~\cite{yates_1934}.

In summary, depending on the shape and statistics of the input data set, the effective number of degrees of freedom of a contingency
table can differ from the advocated value of $n_{dof}$ (Eqn.~\ref{eq:ndof}).
To be certain of the effective value to use, this is best derived as the average value of the test statistic distribution,
which is obtained with Monte Carlo simulations of the expected frequency distribution.

\subsection{Modified $\chi^2$ distribution} \label{sec:chi2mod}

Given a large enough data sample, and given the hypothesis that the observed frequencies result from a random sampling from the distribution of
expected frequencies, the $G$-test statistic can be approximated\footnote{The approximation is obtained with a second-order Taylor
  expansion of the logarithm around 1.} by Pearson's $\chi^2$. 
In this scenario both the $G$-test and $\chi^2$ value are described by the $\chi^2$ distribution of Eqn.~\ref{eq:chi2dist}, with the same number of degrees of freedom,
and applying any one test leads to the same conclusions.

\begin{figure}[htp]
  \centering
  \begin{subfigure}[t]{0.5\linewidth}
    \centering
    \includegraphics[width=\textwidth]{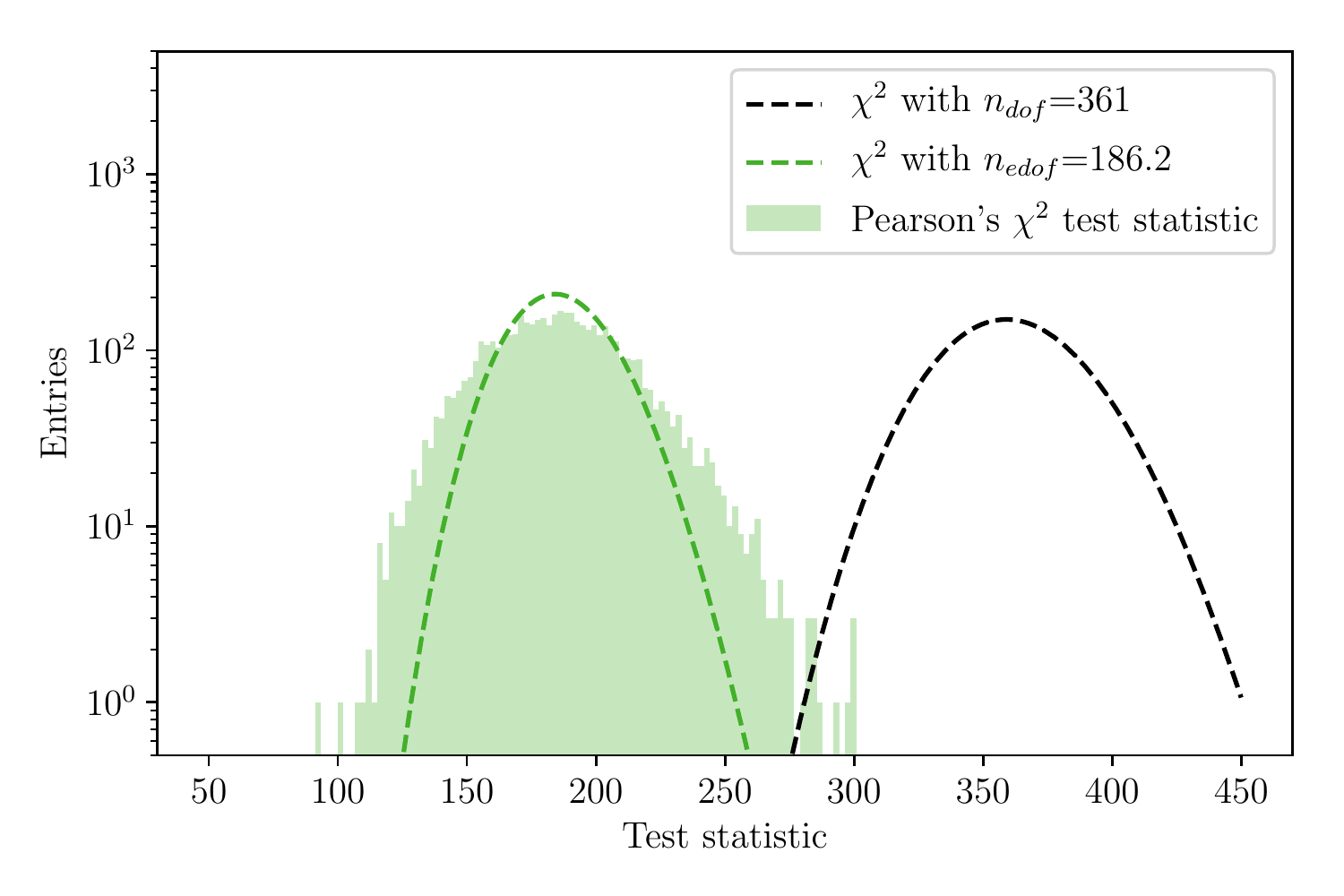}
    \caption{}
  \end{subfigure}%
    \begin{subfigure}[t]{0.5\linewidth}
    \centering
    \includegraphics[width=\textwidth]{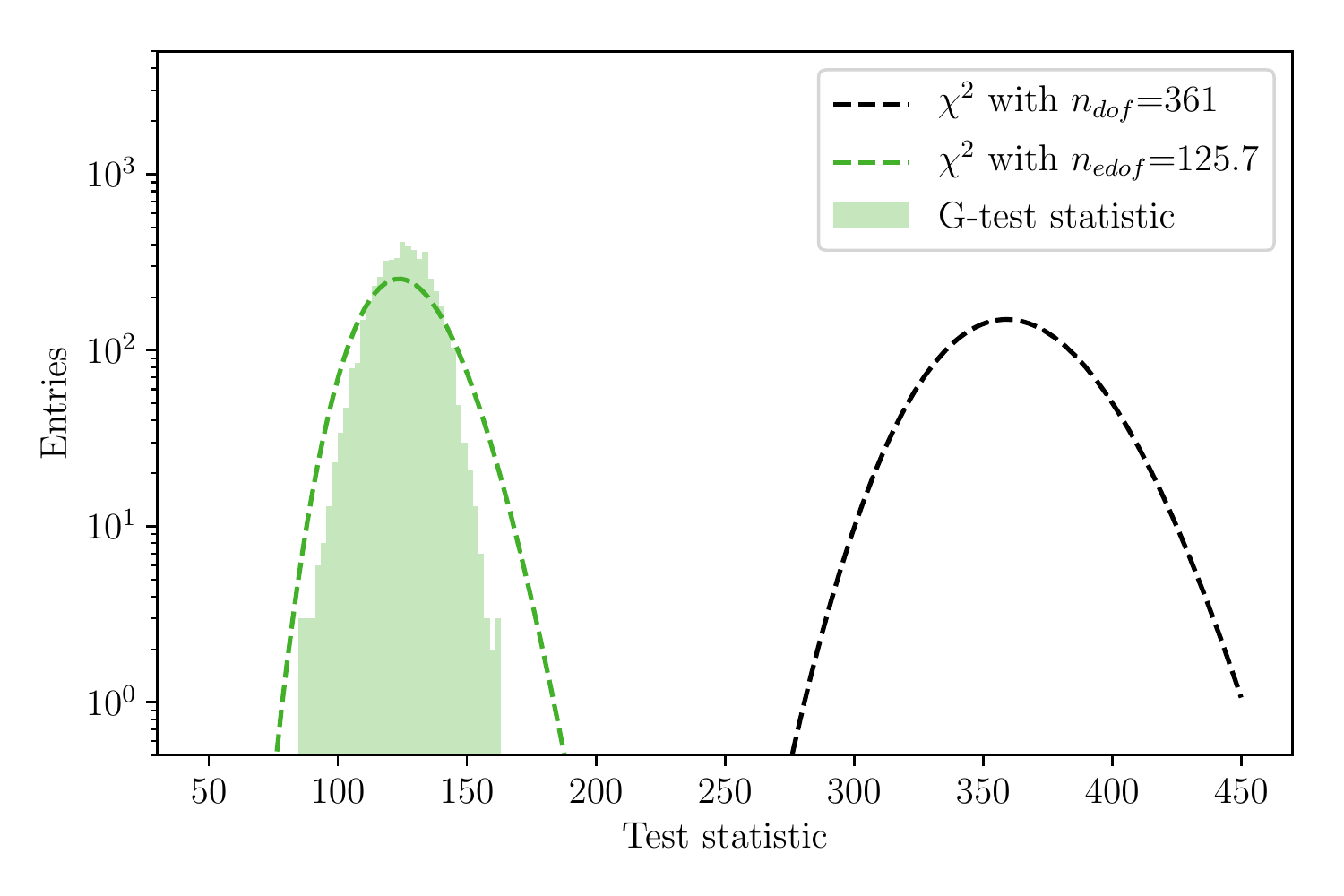}
    \caption{}
  \end{subfigure}\\%
  \begin{subfigure}[t]{0.5\linewidth}
    \centering
    \includegraphics[width=\textwidth]{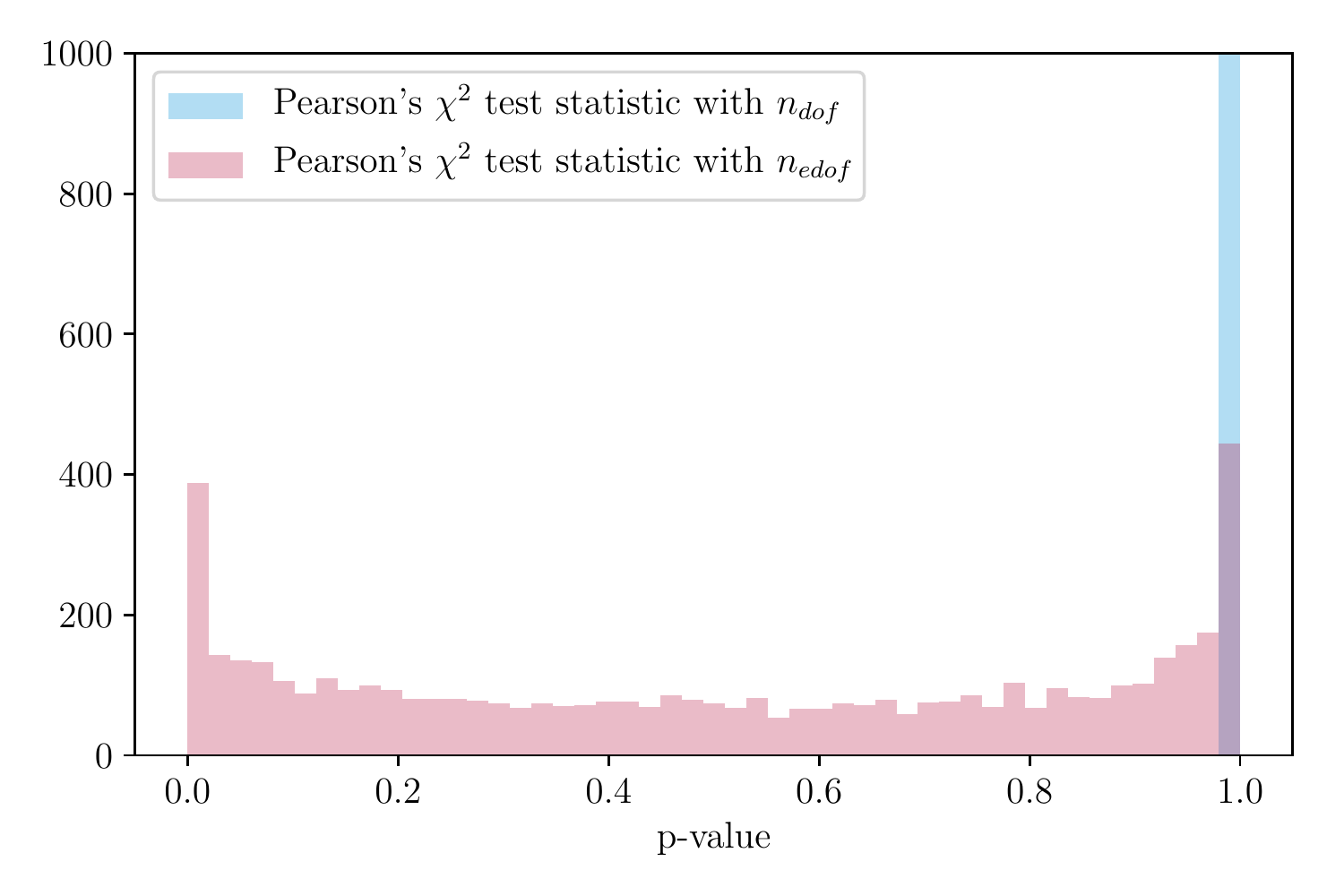}
    \caption{}
  \end{subfigure}%
    \begin{subfigure}[t]{0.5\linewidth}
    \centering
    \includegraphics[width=\textwidth]{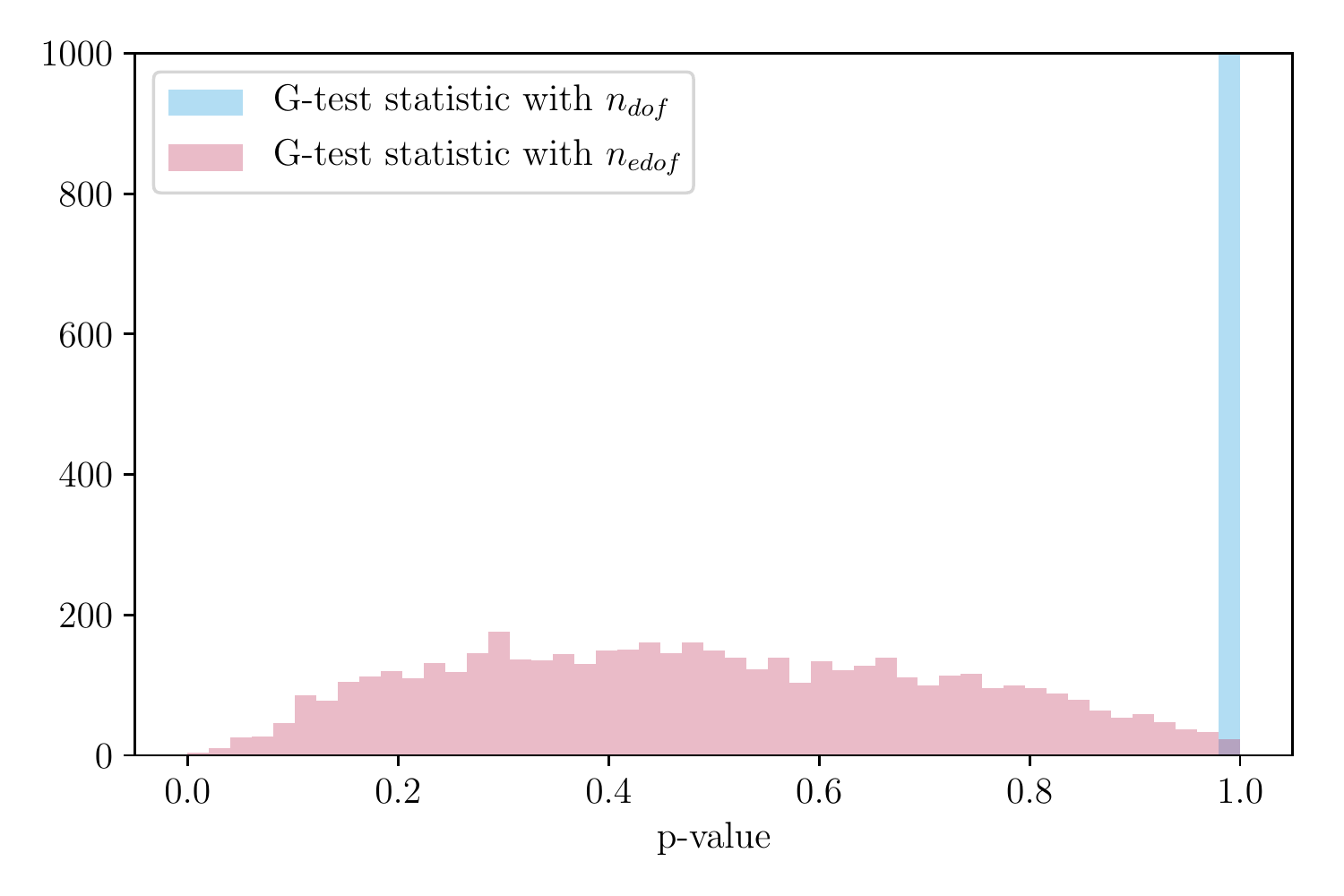}
    \caption{}
  \end{subfigure}%
    \caption{The simulated distribution of $\chi^2$ and $G$-test statistics using the smiley data set as input (Figure~\ref{fig:smiley}).
      a) The $\chi^2$ distribution (green) is wider than the expected distribution (dashed lines), and b) the $G$-test distribution (green) is narrower.
      The corresponding $p$-value distributions are shown in panels c) and d), both when using $n_{dof}$ (blue) and $n_{edof}$ (blue).}
  \label{fig:divergence}
\end{figure}

For low statistics samples -- to be more specific, samples with many bins of low expected and observed frequencies -- the distributions of $G$ and $\chi^2$ start to differ,
and both distributions diverge from the nominal $\chi^2$ distribution. 
This can be seen in Fig.~\ref{fig:divergence}, which uses the smiley data set of Fig.~\ref{fig:smiley} as input.
The simulated distribution of test statistics is wider than the $\chi^2$-distribution in case of the Pearson $\chi^2$-test statistic (Fig.~\ref{fig:divergence}a) and narrower than the $\chi^2$-distribution in case of the $G$-test statistic (Fig.~\ref{fig:divergence}b).
This results in $p$-value distributions with elevated frequencies around zero and one for the Pearson $\chi^2$-test statistic (Fig.~\ref{fig:divergence}c) and lower frequencies near zero and one for the $G$-test statistic (Fig.~\ref{fig:divergence}d).
Note that the effective number of degrees of freedom is much lower than the theoretical value; using $n_{\rm dof}$ in the $p$-value calculation results in an uneven distributions peaked towards one.

This section addresses the question whether the test statistic distribution for the contingency test can be modeled for all sample sizes,
knowing that Eqn.~\ref{eq:chi2dist} cannot be safely used for low statistics data sets.
In particular we are interested in assessing the $p$-values of large test statistic values, coming from possibly strong variable dependencies.
To evaluate these correctly, it is important to properly model the high-end tail of the test statistic distribution.

We observe empirically that for low-statistics samples the $G$-test statistic distribution converges towards a Gaussian
distribution $G(x|\mu,\sigma)$, with mean $\mu=n_{\rm edof}$ and width $\sigma=\sqrt{n_{\rm edof}}$.
For high-statistics samples the distribution is modeled by $g(x|k)$, with $k=n_{\rm edof}$ degrees of freedom.
Experimentally we find that, for any sample size, the $G$-test statistic distribution can be well
described by the combined probability density function $h(x|f)$:
\begin{equation} \label{eq:chi2mod}
h(x|f) = f \cdot g(x|n_{\rm edof}) + (1-f)\cdot G(x|n_{\rm edof},\sqrt{n_{\rm edof}})\,,
\end{equation}
where the parameters of $g(x|k)$ and $G(x|\mu,\sigma)$ are fixed as above, and $f$ is a floating fraction parameter between $[0,1]$.

Below we use $h(x|f)$ as the modified $\chi^2$ p.d.f. to model the $G$-test statistic distribution for any data set.

\begin{figure}[htp]
  \centering
  \begin{subfigure}[t]{0.5\linewidth}
    \centering
    \includegraphics[width=\textwidth]{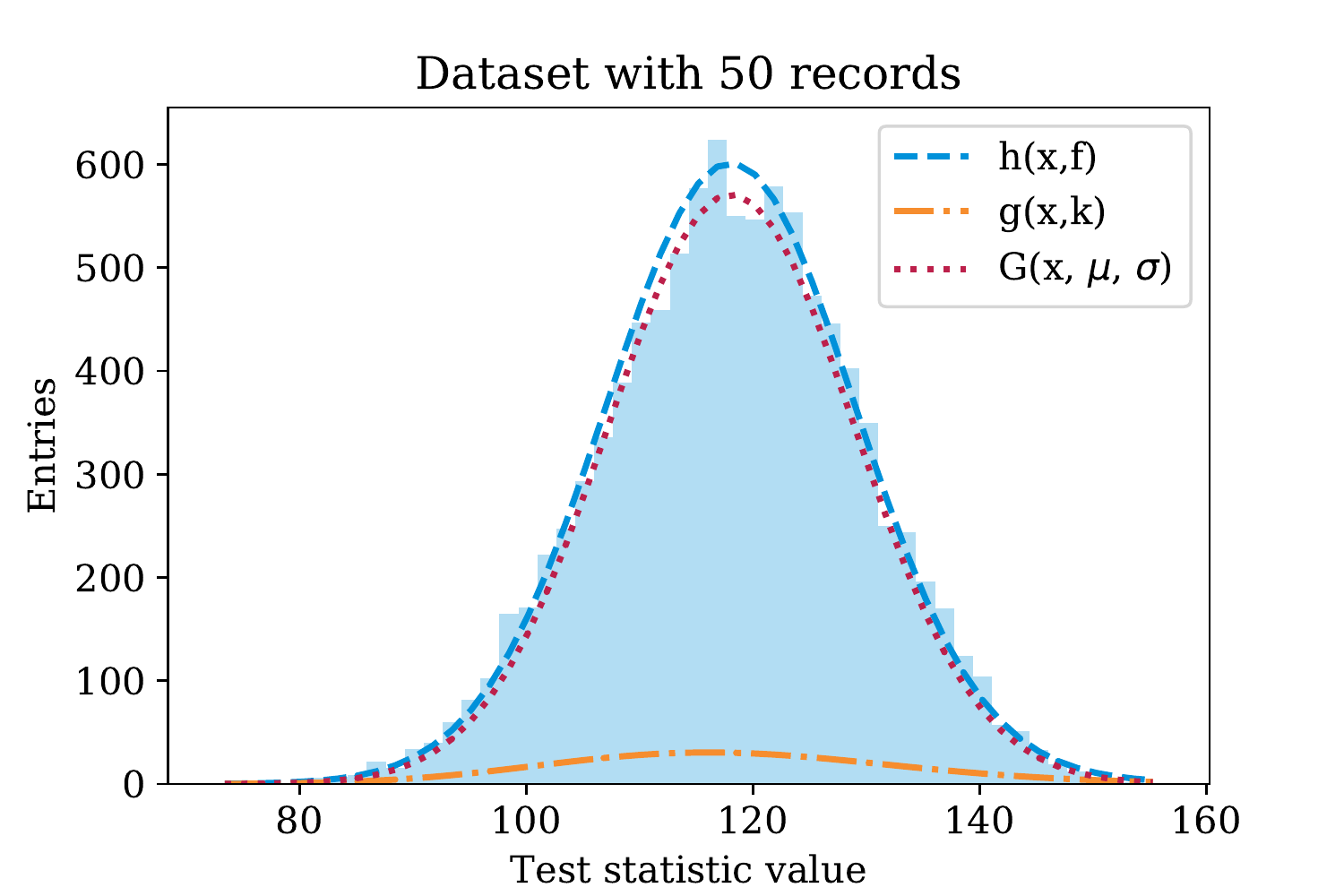}
    \caption{}
  \end{subfigure}%
  \begin{subfigure}[t]{0.5\linewidth}
    \centering
    \includegraphics[width=\textwidth]{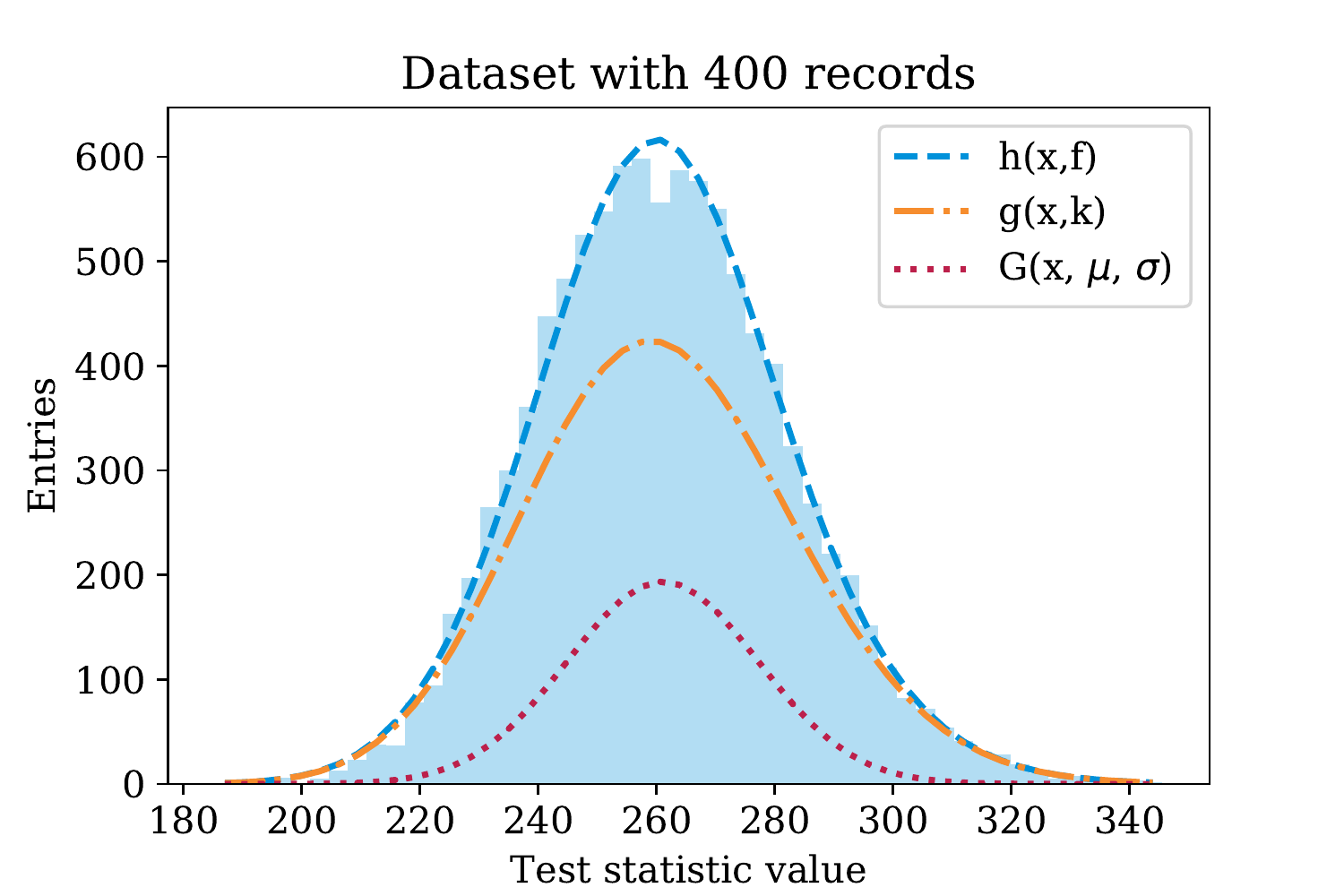}
    \caption{}
  \end{subfigure}%
  \caption{The $G$-test statistic distribution for two smiley data sets containing a) 50 and b) 400 data points. The distribution is modeled with the $h(x|f)$ distribution. }
  \label{fig:gtestdistfit}
\end{figure}

Fig.~\ref{fig:gtestdistfit} shows the results of binned log-likelihood fits of $h(x|f)$ to two $G$-test statistic distributions,
each with 10k entries generated with the procedure of Section~\ref{sec:nedof}, using the smiley data set with 20x20 bins
with: a) $N=50$ and b) $N=400$ records for the simulated data sets.
Clearly, these distributions are not well modeled using $g(x|n_{\rm edof})$ or $G(x|n_{\rm edof},\sqrt{n_{\rm edof}})$ alone.
The fit of $h(x|f)$ can separate the two component p.d.f.'s given that
the RMS-value of $g(x|n_{\rm edof})$ is $\sqrt{2n_{\rm edof}}$
and the width of the Gaussian is fixed to $\sqrt{n_{\rm edof}}$.
For $N=50$, the distribution is dominated by the Gaussian, and for $N=400$ by the theoretical $\chi^2$ distribution.
Note that $G(x|n_{\rm edof},\sqrt{n_{\rm edof}})$, when present, contributes to the core of the distribution while $g(x|n_{\rm edof})$ dominates in the tails.

\begin{figure}[htp]
  \centering
  \begin{subfigure}[t]{0.5\linewidth}
    \centering
    \includegraphics[width=\textwidth]{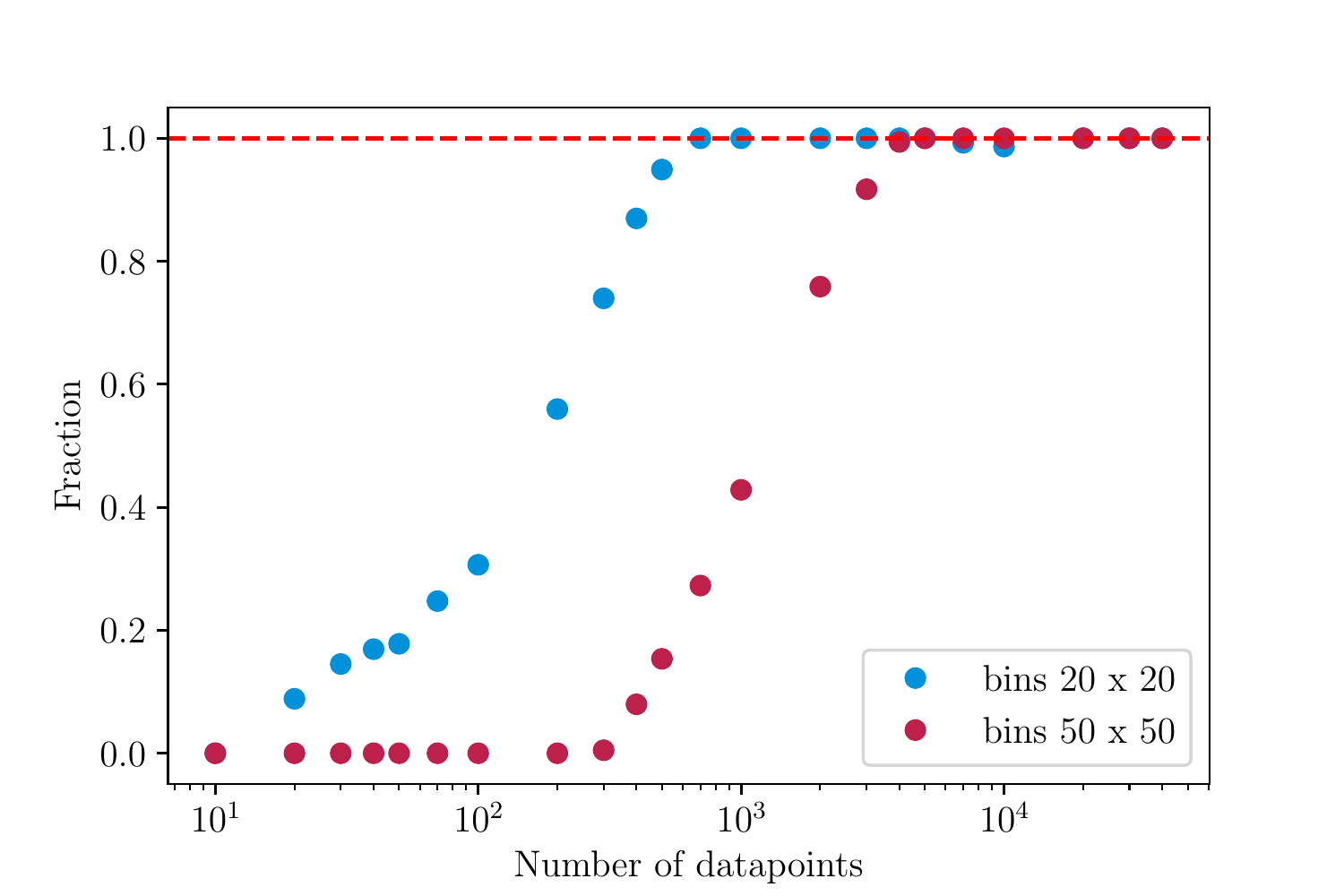}
    \caption{}
  \end{subfigure}%
  \begin{subfigure}[t]{0.5\linewidth}
    \centering
    \includegraphics[width=\textwidth]{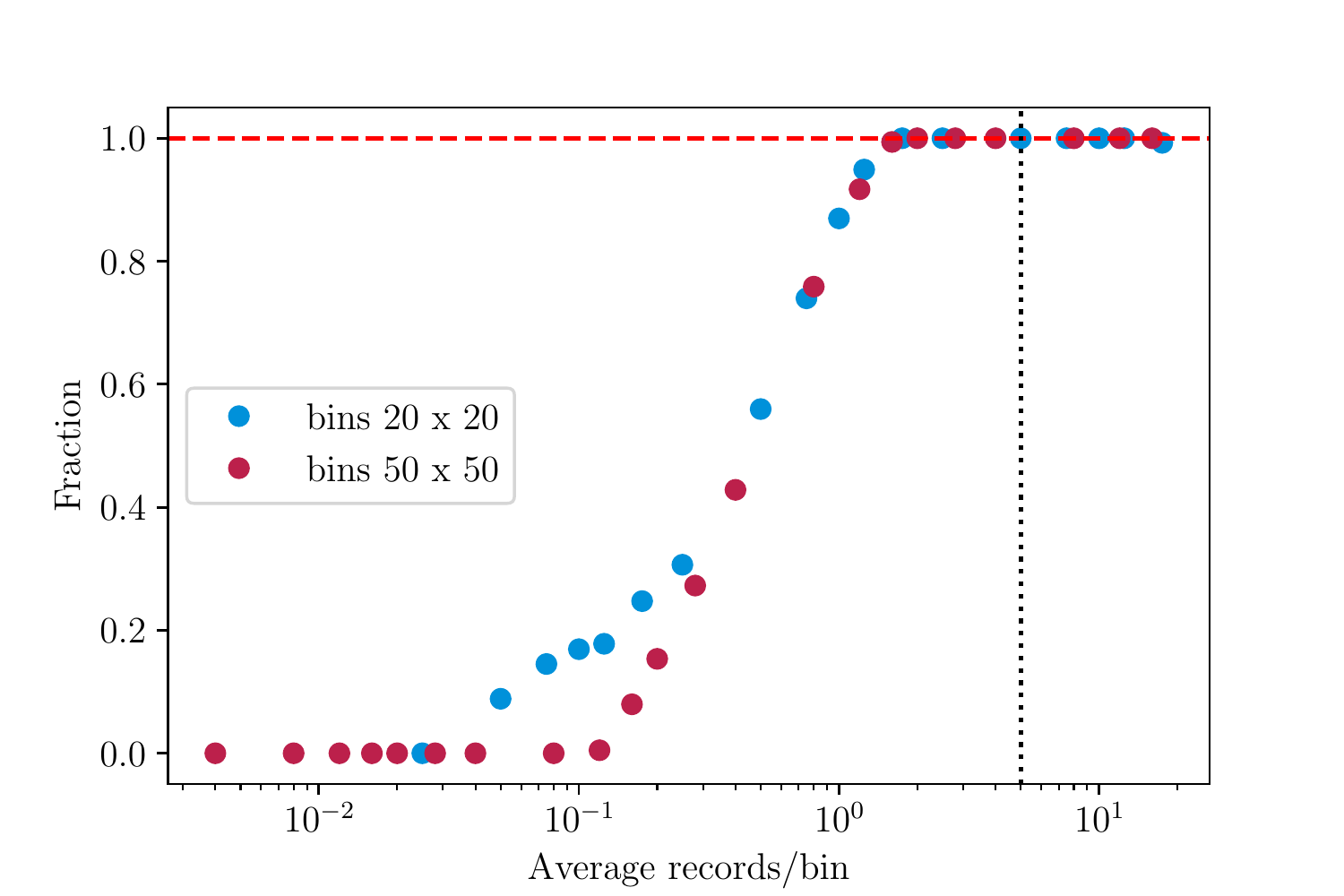}
    \caption{}
  \end{subfigure}%
  \caption{a) The fit fraction $f$ as a function of the number of records per simulated data set, $N$. b) The same data points, but here $f$ is shown
    as a function of the average number of records per bin.}
  \label{fig:ffitrize}
\end{figure}

Fig.~\ref{fig:ffitrize} uses a similar setup, with 20x20 or 50x50 bins, where the fit fraction $f$ is shown as a function of a) the number
of records per simulated data set, $N$, and b) the average number of records per cell, $\bar{n}$.
The fraction $f$ rises as a function of sample size, such that $h(x|f)$ turns into $g(x|n_{\rm edof})$ for large enough data sets.
With 20x20 bins, for a fraction of $0.50$ ($0.99$) the approximately sample size equals $175$ ($700$),
and the average number of entries per cell equals $0.4$ ($1.8$).
Note that the fraction reaches $1$ well before $n_{\rm edof}$ reaches the advocated value of $n_{\rm dof}$ in Fig.~\ref{fig:nedof}.

In summary, to assess the $p$-value for the hypothesis test of variable independence,
in this work we choose to work with the $G$-test statistic, and not Pearsons's $\chi^2$, for two reasons:
\begin{enumerate}
\item We manage to describe the $G$-test statistic distribution most successfully for any sample size.
\item As seen from Fig.~\ref{fig:divergence}b, for a large observed test statistic value, corresponding to a large significance of variable dependency,
  applying the naive formula of Eqn.~\ref{eq:pvalue} over-covers, \textit{i.e.} gives a conservative $p$-value (the green distribution is narrower than expected).
\end{enumerate}

We use the distribution $h(x|f)$ of Eqn.~\ref{eq:chi2mod} as modified $\chi^2$ distribution in Eqn.~\ref{eq:pvalue}
to assess the $p$-value for the hypothesis test.

\subsection{Evaluation of significance} \label{sec:signfeval}

The statistical significance of the hypothesis test of any variable independence is obtained with the following procedure:
\begin{enumerate}
\item Calculate the average number of entries per cell, $\bar{n}$.
  If $\bar{n} < 4$, set $n_{\rm sim} = 2000$, else $n_{\rm sim} = 500$ samples.
\item Follow the procedure of Section~\ref{sec:nedof} to generate $n_{\rm sim}$ synthetic data sets based on the dependent frequency estimates of the input data set.
  For each synthetic data set evaluate its $G$-test value. Take the average of the $G$-test distribution to obtain $n_{\rm edof}$.
\item If $\bar{n} < 4$, to obtain $f$ fit the probability density function $h(x|f)$
  to the $G$-test distribution, with $n_{\rm edof}$ fixed.
  Else, skip the fit and set $f=1$.
\item With this fraction, use Eqn.~\ref{eq:pvalue} with $h(x|f)$ as modified $\chi^2$ distribution to obtain the $p$-value for the hypothesis test,
  using the $G$-test value from data as input.
\item The $p$-value is converted to a normal $Z$-score:
\begin{equation} \label{eq:zscore}
Z = \Phi^{-1}(1-p)\ ;\quad \Phi(z)=\frac{1}{\sqrt{2\pi}} \int_{-\infty}^{z} e^{-t^{2}/2}\,{\rm d}t\,,
\end{equation}
where $\Phi^{-1}$ is the quantile (inverse of the cumulative distribution) of the standard Gaussian,
\textit{e.g.} $Z$ is the significance in 1-sided Gaussian standard deviations.
For example, the threshold $p$-value of $0.05$ (95\% confidence level) corresponds to $Z=1.64$.

When the $p$-value is too small to evaluate Eqn.~\ref{eq:zscore} numerically, at $p\lesssim 10^{-310}$, anyhow a very strong variable dependency,
$Z$ is estimated using Chernoff's bound~\cite{lin_genest_banks_molenberghs_scott_wang_2014} to ensure a finite value.
Let $z\equiv G/n_{\rm edof}$, Chernoff states when $z>1$: 
\begin{equation}
p \leq f \cdot (z e^{1-z})^{n_{\rm edof} / 2}\,,
\end{equation}
where we safely ignore the contribution from the narrow Gaussian in $h(x|f)$.
This is converted to $Z$ with the approximation (valid for large $Z>1.5$):
\begin{equation} 
Z = \sqrt{u- \log{u}} ;\quad u = -2\log{(p \sqrt{2\pi})}\,.
\end{equation}
\end{enumerate}

\begin{figure}[htp]
  \centering

  \begin{subfigure}[t]{0.5\linewidth}
    \centering
    \includegraphics[width=\textwidth]{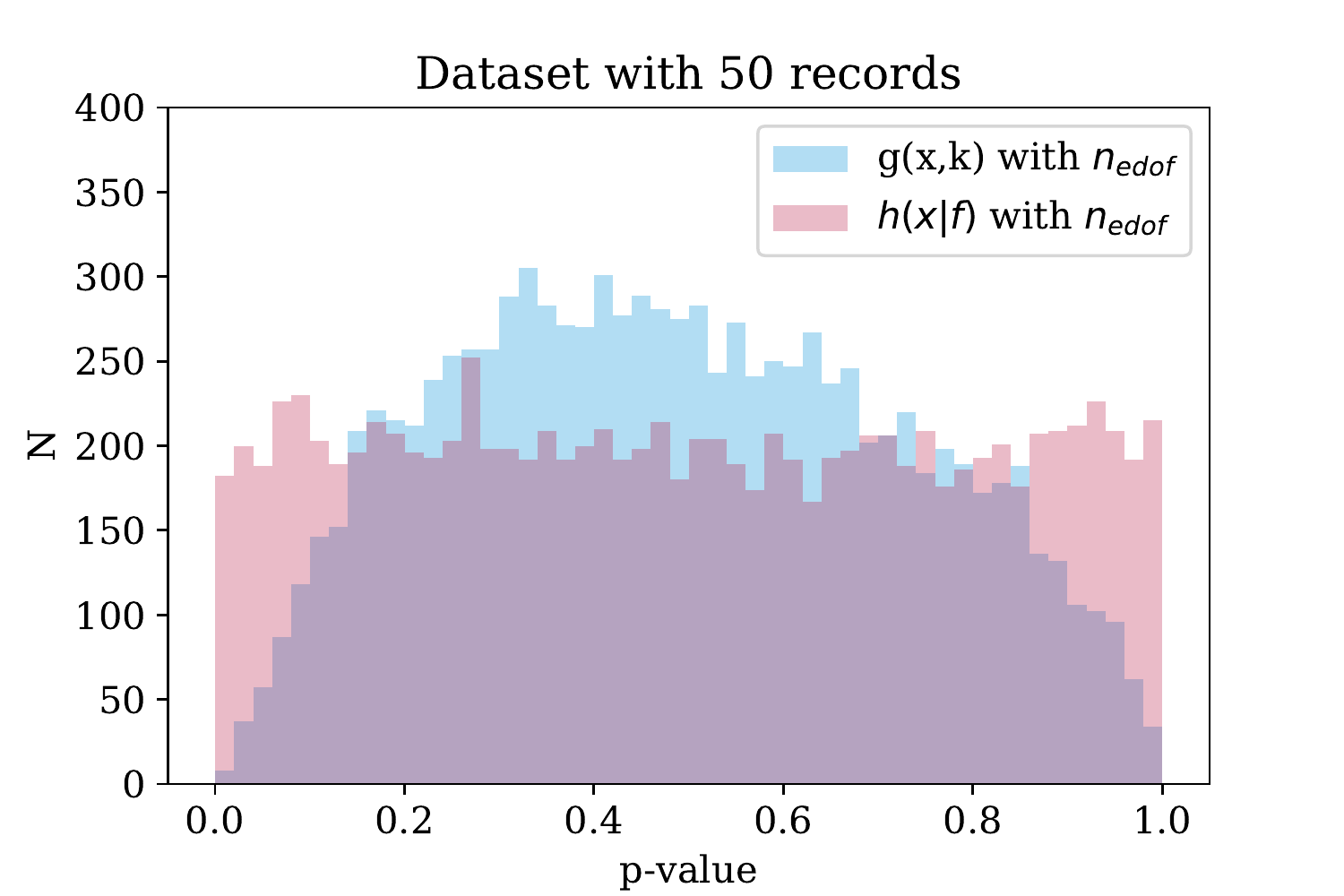}
    \caption{}
  \end{subfigure}%
  \begin{subfigure}[t]{0.5\linewidth}
    \centering
    \includegraphics[width=\textwidth]{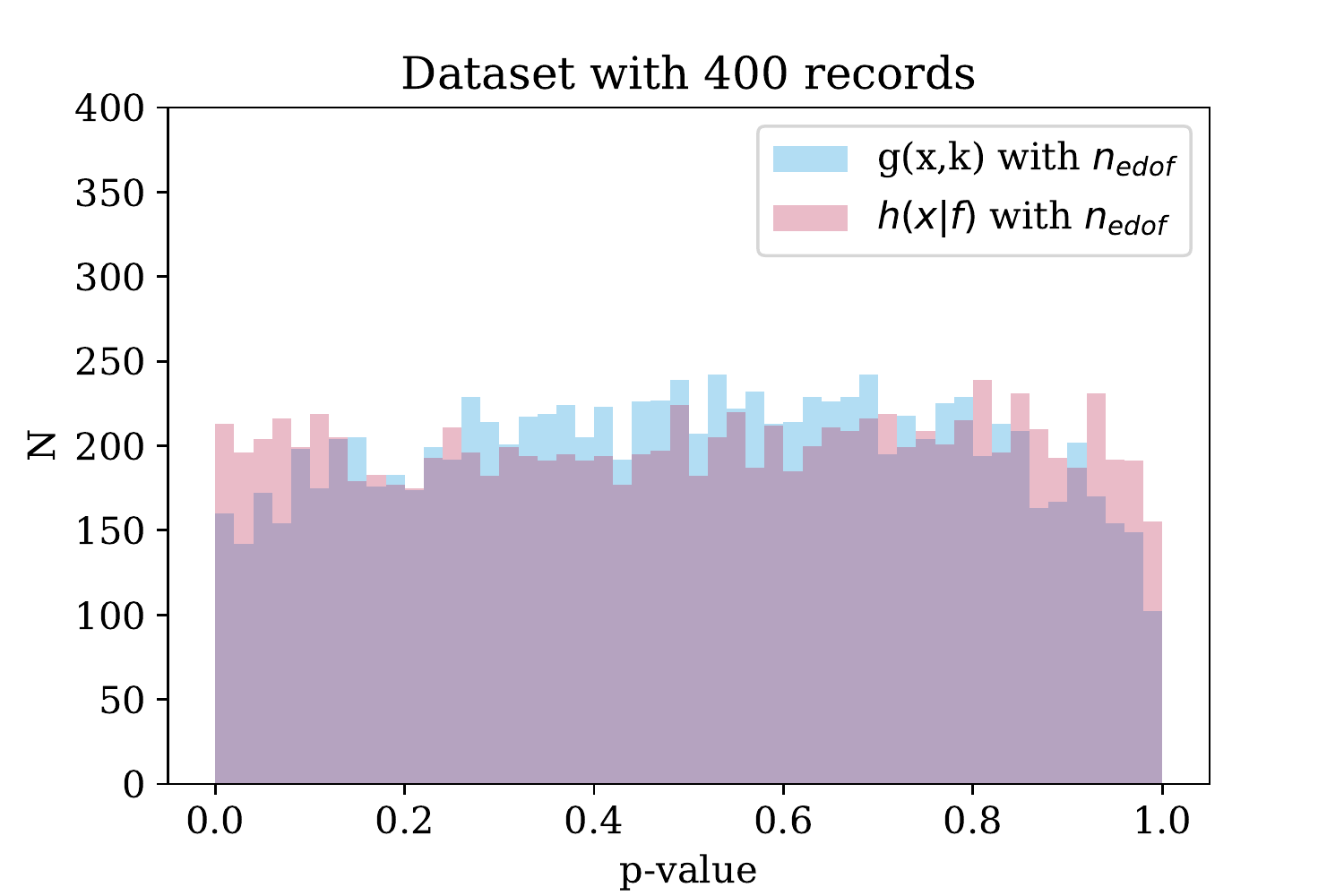}
    \caption{}
  \end{subfigure}%
  \caption{The $p$-value distributions corresponding of the two $G$-test distributions of Fig.~\ref{fig:gtestdistfit}, with a) $N=50$ and b) $N=400$ records per sample.
    See the text for a description of the two $p$-value calculations performed.}
  \label{fig:significance}
\end{figure}

The significance procedure is illustrated in Fig.~\ref{fig:significance}, which shows the $p$-value distributions of the two
$G$-test distributions of Fig.~\ref{fig:gtestdistfit}, with $N=50$ and $N=400$ records per sample.
The two $p$-value distributions in each figure have been calculated in two ways. 
\begin{enumerate}
  \item Using the original $\chi^2$ distribution $g(x|k)$ of Eqn.~\ref{eq:pvalue}, with the effective number of degrees of freedom, $n_{\rm edof}$.
This results in the blue distributions.
\item Fitting each test statistic distribution with $h(x|f)$ of Eqn.~\ref{eq:chi2mod}, and using that to calculate the $p$-values,
  resulting in the red distributions.
\end{enumerate}
The blue distributions drop around zero and one, in particular for the low statistics sample ($N=50$).
This is because the $G$-test distribution is more narrow than the $\chi^2$ distribution, as shown in Fig.~\ref{fig:divergence}.
The red $p$-value distributions, evaluated with $h(x|f)$, are uniform, as desired in both setups.

Let us apply the statistical procedure to a low-statistics data sample.
A smiley data set with 100 entries, in a histogram with 20x20 bins, has correlation value $\phik=0.73$ and test statistic value $G = 227.4$.
The $Z$ calculation is done in three consecutively more refined ways:
\begin{enumerate}
\item \textit{The asymptotic approximation}: using $n_{\rm dof} = 361$ and the asymptotic $\chi^2$ distribution $g(x|k)$ gives: $Z = -5.7$;
\item \textit{Effective number of degrees of freedom}: using $n_{\rm edof} = 189.3$ and the asymptotic $\chi^2$ distribution $g(x|k)$ results in: $Z = 1.9$;
\item \textit{Modified $\chi^2$ distribution}: with $n_{\rm edof} = 189.3$, the modified $\chi^2$ distribution $h(x|f)$, and fit fraction $f = 0.10$ one finds: $Z = 2.5$.
\end{enumerate}
In this example, between the three approaches the $Z$-value increases with more than 8 units!
Typically, using the effective number of degrees of freedom gives the largest correction to $Z$, and
the modified $\chi^2$ distribution only gives a small correction on top of that.

The choice of 2000 synthetic data sets for the fit of $h(x|f)$ is a compromise between accuracy and speed.
With this number, $Z$ typically varies at the level of $0.04$, and is calculated in just a fraction of a second.

Based on our findings, for any sample size we recommend 
the $p$-value to be calculated with the modified $\chi^2$ distribution $h(x|f)$, using $n_{\rm edof}$ degrees of freedom.
If not, the $p$-value may over-cover for strong variable dependencies and at low-statistics,
resulting in a $Z$-value that is too small, possibly by multiple units.
This is important to know, as it can lead to rather incorrect conclusions regarding the studied variable dependency.

\subsection{Example significance matrix}

In practice a correlation value may be small but its statistical significance can still be large, and vice versa.
For this reason, when exploring a data set, the levels of correlation and significance should always be studied together.

Fig.~\ref{fig:significance_matrix} shows the significance matrix determined for the car insurance data set of Table~\ref{tab:data}.
Compared with the correlation matrix of Fig.~\ref{fig:phik_example}, the low \phik values happen to be statistically insignificant,
but the higher values are very significant.

\begin{figure}[htp]
  \centering
  \begin{minipage}[b]{0.6\linewidth}
    \centering
    \includegraphics[width=\textwidth]{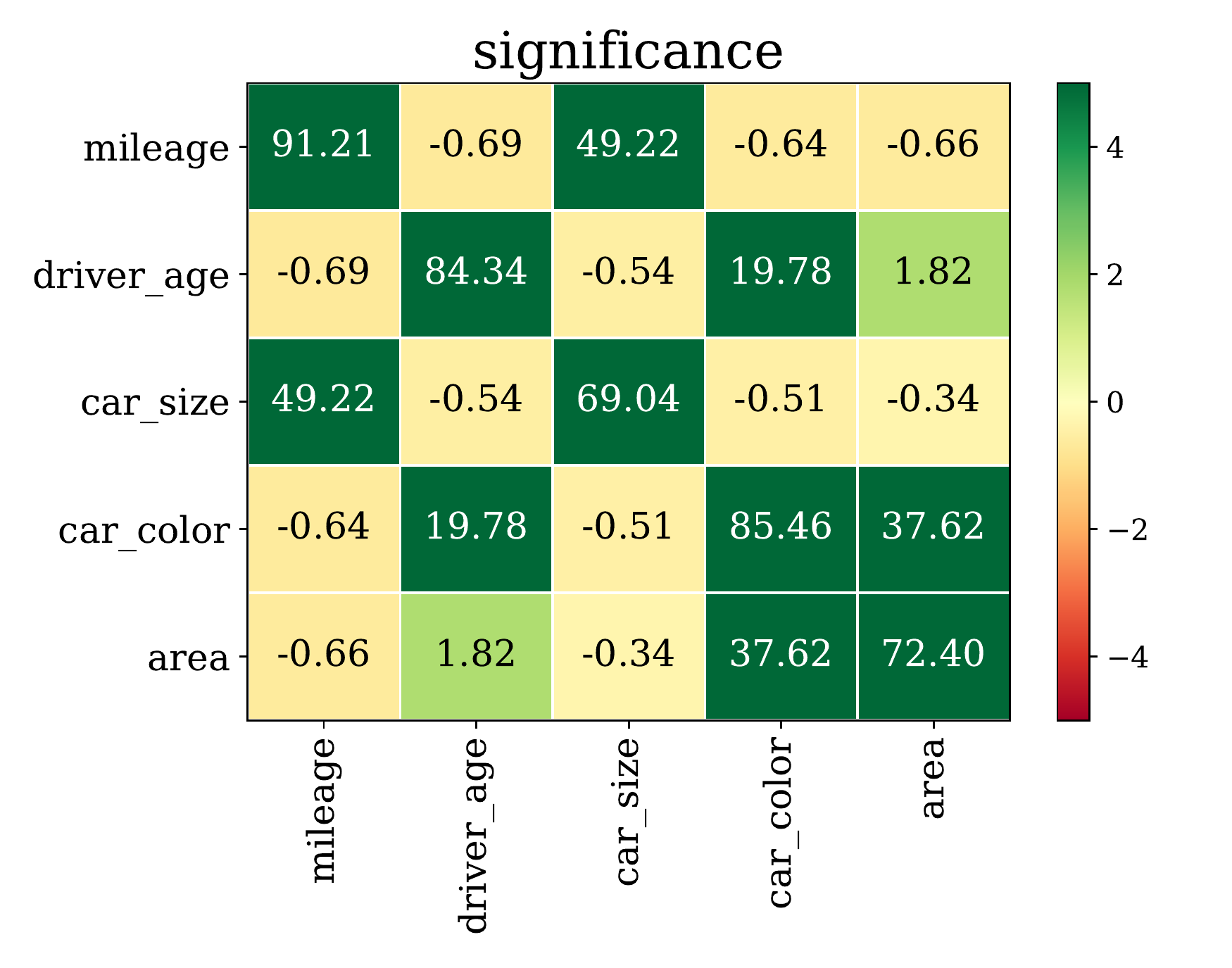}
  \end{minipage}%
  \caption{The significance matrix, showing the statistical significances of correlated and uncorrelated variable pairs.
    The color scale indicates the level of significance, and saturates at $\pm5$ standard deviations.}
  \label{fig:significance_matrix}
\end{figure}

\subsection{Sampling approaches} \label{sec:simapp}

Based on the statistically dependent frequency estimates,
three sampling approaches are offered to generate synthetic data sets for testing the hypothesis of no variable association:
\begin{itemize}
\item \textit{Multinomial sampling}: with only the total number of records fixed. The hypothesis of no association is independent of the row and column variables.
\item \textit{Product-multinomial sampling}: with the row or column totals fixed in the sampling. The hypothesis of no association is also called homogeneity of proportions.
  This approach is commonly used in cohort and case-control studies.
\item \textit{Hypergeometric sampling}: both the row or column totals are fixed in the sampling. This approach is also known as Fisher's exact test.
  We use Patefield's algorithm~\cite{patefield_1981} to generate the samples. 
\end{itemize}

There is an ongoing debate about sampling design for tests of variable independence.
Although in practice most people are not too worried about the sampling approach, at least not in the high-statistics regime,
because asymptotically the different approaches lead to the same result. 
The default approach used in this paper is multinomial sampling.
For a discussion and further references see Ref.~\cite{kim_agresti_1997}.

\section{Interpretation of relation between two variables}
\label{sec:outliers}

After the evaluation of \phik and its significance, the specific relationship between two variables is typically inspected.
To facilitate the interpretation of any dependency found, 
the significance of observed excesses or deficits of records with respect to expected values in the contingency table is discussed here.
%

The statistical significance for each cell in the table is obtained from an hypothesis test between a background-only
and signal-plus-background hypothesis for a Poisson process.
Such hypothesis tests, \textit{i.e.} for the presence of new sources of (Poisson) counts on top of known ``background'' processes, are frequently performed in
many branches of science, for example gamma ray astronomy and high energy physics,
and have been discussed extensively in the literature~\cite{Cousins:2007bmb}. 

We employ a measure of statistical significance commonly used in both fields, one that accounts for the
mean background rate having a non-negligible uncertainty.
The background estimate and its uncertainty have been derived from an auxiliary or side-band measurement,
typically assumed to be a Poisson counting setup, as in the case of the ABCD estimate of Section~\ref{sec:indepfreq}.
Here we use as background estimate the statistically independent frequency estimate (and related uncertainty) of Eqn.~\ref{eq:abcd} (\ref{eq:abcderror}).

The hybrid Bayesian-Frequentist method from Linneman~\cite{Linnemann:2003vw} 
is used to evaluate the probability of the hypothesis test ($p$-value).
Per cell, Linneman's probability calculation requires
the observed count $n_o$, the expected count $n_e$, and the uncertainty on the expectation $\sigma_e$:
\begin{equation} \label{eq:linneman}
p_{B} = B \big( 1/(1+\tau),\, n_o,\, n_e\, \tau + 1 \big) \,,
\end{equation}
where $B$ is the incomplete Beta function, and $\tau = n_e / \sigma_e^2$.

We apply four corrections on top of this calculation:
\begin{enumerate}
\item The incomplete Beta function returns no number for $n_o=0$, when by construction the $p$-value should be $1$.
\item The incomplete Beta function is undefined when $\sigma_e=0$, in which case we simply revert to the standard Poisson distribution.
\item The incomplete Beta function always returns $1$ when $n_e=0$, irrespective of $n_o$ and $\sigma_e$. The scenarios $n_o=0$ and $\sigma_e=0$ are
  captured by the previous two fixes. In all other cases we set $n_e=\sigma_e$ before evaluating Eqn.~\ref{eq:linneman}.
  In particular, this procedure prevents (minus) infinite significances for low statistics cells where uncertainty-wise these are not expected\footnote{When $n_e=0$, $B$ or $C$ is zero in Eqn.~\ref{eq:abcd},
  so Eqn.~\ref{eq:abcderror} typically gives $\sigma_e<1$. For example, for $n_e = 0$, $\sigma_e =0.14$, and $n_o=0$ ($1$), correction three to $p_B$ results in $Z = -0.29$ $(1.10)$.
  Varying $n_e$ between $\sigma_e/2$ and $3\sigma_e/2$ gives a maximum absolute shift in $Z$ of $0.05$ ($0.12$). To do outlier detection, for this procedure we deem this level of systematic
  error acceptable.}.
\item As we combine an integer-valued measurement (namely the observed frequency) with a continuous expectation frequency and uncertainty,
resulting in a continuous (combined) test statistic,
we correct $p_B$ to Lancaster's mid-$P$ value~\cite{lancaster_1961}, which is the null probability of more extreme results plus only half the probability of
the observed result\footnote{The standard $p$-value definition is: $p = P(s \geq {\rm observed}\, |\, {\rm background} )$.}: 
\begin{equation}
p = P(s = {\rm observed}\, |\, {\rm background} ) / 2 + P(s > {\rm observed}\, |\, {\rm background} )\,,
\end{equation}
with $s$ the integrated-over number of cell counts.
This $p$-value is then translated into the $Z$-value using Eqn.~\ref{eq:zscore}.
When observing the expected frequency by construction Lancaster's mid-$P$ value ($Z$-value) is close to $0.5$ ($0$), even at low statistics.
Likewise, for background-only samples the Lancaster's mid-$P$ correction centers the $Z$ distribution around zero.
\end{enumerate}

\begin{figure}[htp]
  \centering
  \begin{subfigure}[t]{0.45\linewidth}
    \centering
    \includegraphics[width=\textwidth]{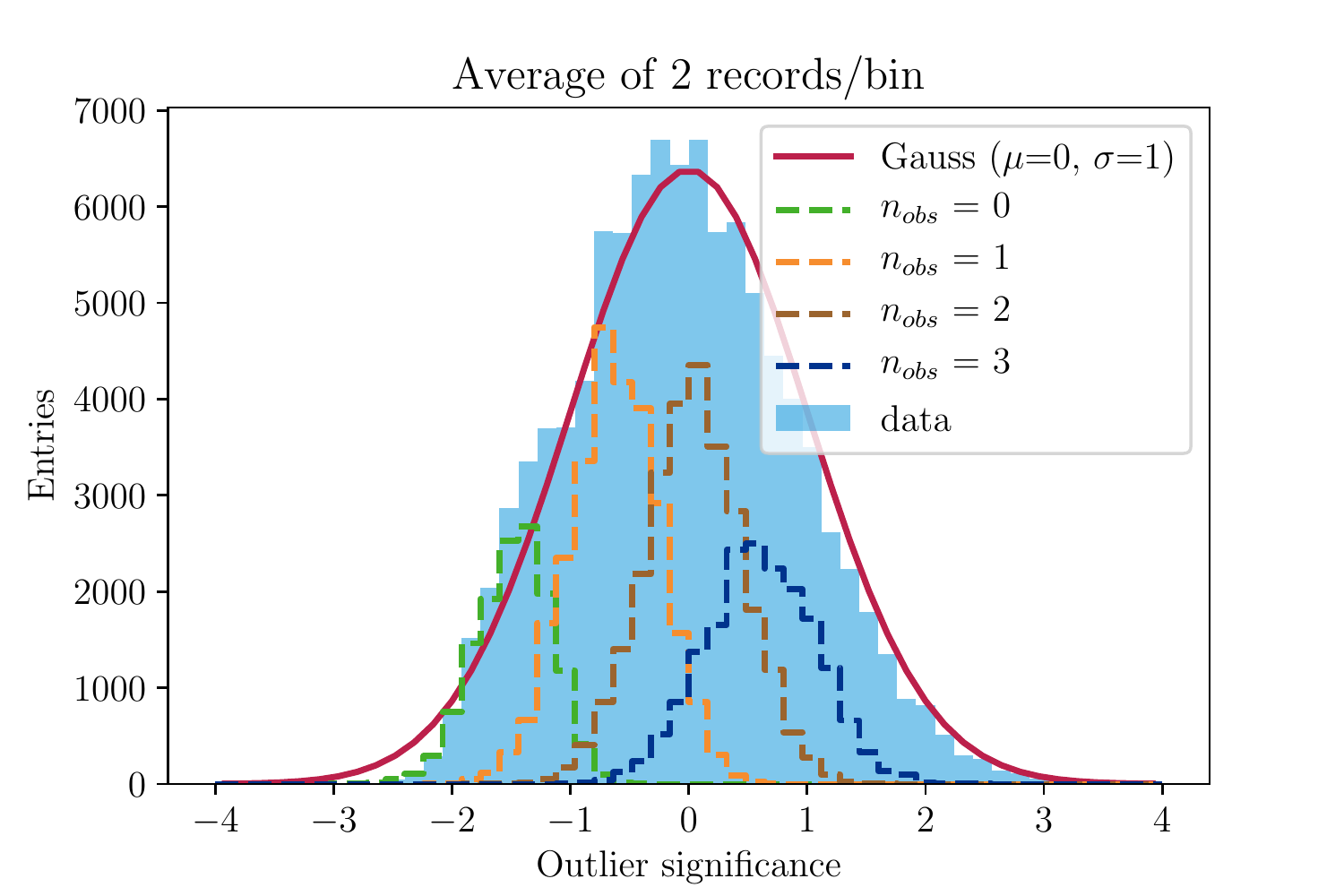}
    \caption{}
  \end{subfigure}
    \begin{subfigure}[t]{0.45\linewidth}
    \centering
    \includegraphics[width=\textwidth]{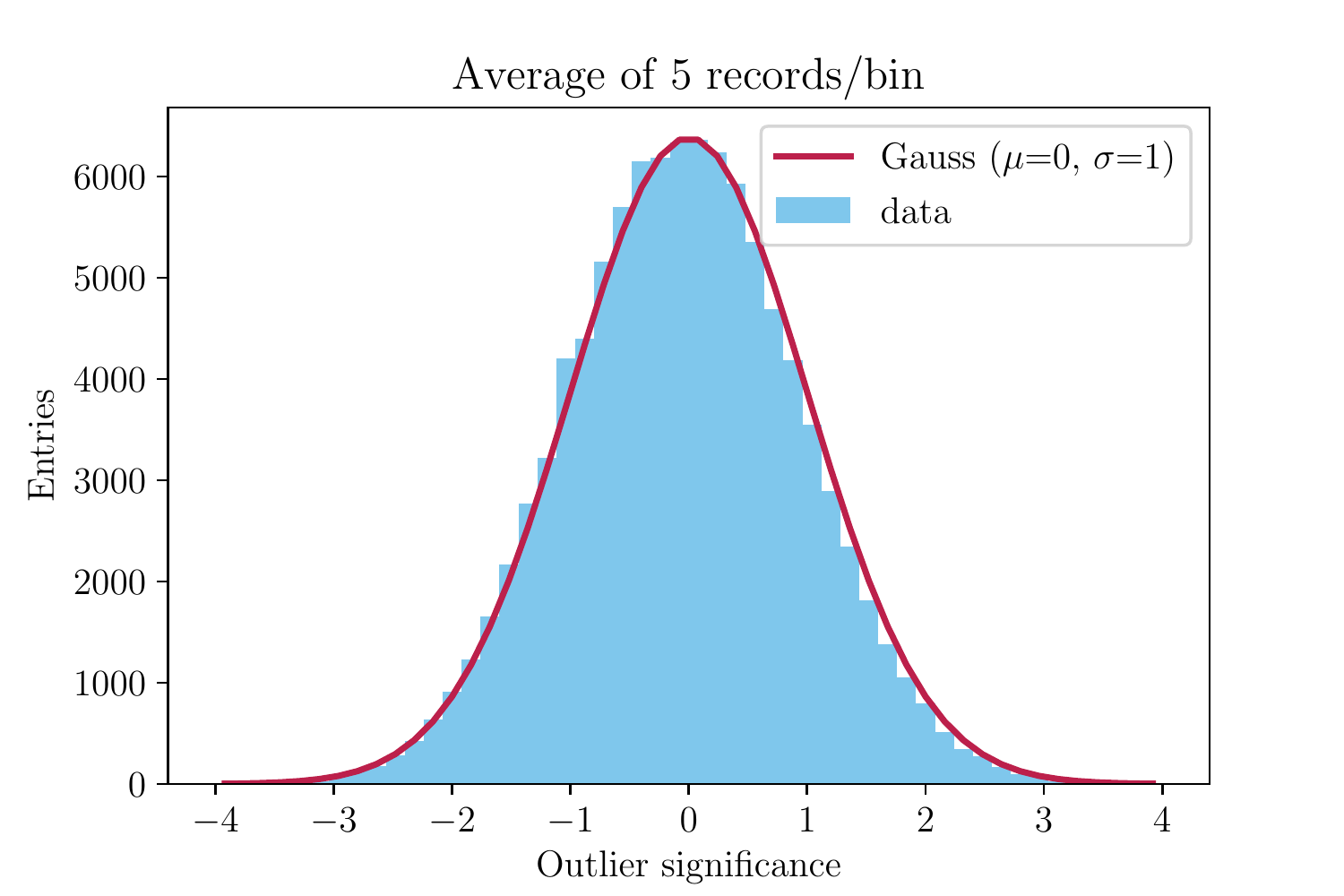}
    \caption{}
  \end{subfigure}%
    \caption{The distribution of outlier significances measured in 1000 randomly generated data sets of two variables obeying a uniform probability mass distribution
      for a data set containing a) 200 and b) 500 records, collected in a $10\times 10$ contingency table. Normal distributions have been overlaid.
      In plot a) the $Z$ distributions from $0$, $1$, $2$, and $3$ observed entries per cell are shown as well.}
  \label{fig:outlierz}
\end{figure}

Fig.~\ref{fig:outlierz}a shows the $Z$ distribution from 1000 randomly generated samples of two variables obeying a uniform probability mass distribution,
\textit{i.e.} the samples have no variable dependency.
Each sample contains only 200 records collected in a $10\times 10$ contingency table,
so on average each cell contains $2.0$ records.
As can be seen from the Gaussian curve, even for such low statistics samples the $Z$ distribution is
fairly consistent with a normal distribution, albeit slightly shifted towards negative values.
Fig.~\ref{fig:outlierz}b shows a similar distribution, built from samples with on average $5.0$ records per contingency table cell.
Clearly, with more statistics the distribution converges to the normal distribution relatively quickly\footnote{With an average
  of less than $1.0$ records per bin, the $Z$ distribution gets more distorted,
  and breaks up into individual peaks of $0$, $1$, $2$, etc. observed entries per cell.
  The distribution peaks at negative $Z$ values, corresponding to no observations, and the tail at negative $Z$ gets truncated.
  Relevant here: the mean of the distribution remains close to zero,
  its width is (slightly less than) one, and the positive tail is similar to that of a normal distribution.}.

To filter out significant excesses or deficits of records over expected values in the contingency table,
one simply demands $|Z|$ to be greater than a specified value, \textit{e.g.} 5 standard deviations.
For two variables with a dependency, note that excesses and deficits always show up together,
since the frequency estimates of Section~\ref{sec:indepfreq} smooth the input distribution.

\begin{figure}[htp]
  \centering
  \begin{subfigure}[t]{0.49\linewidth}
    \centering
    \includegraphics[width=\textwidth]{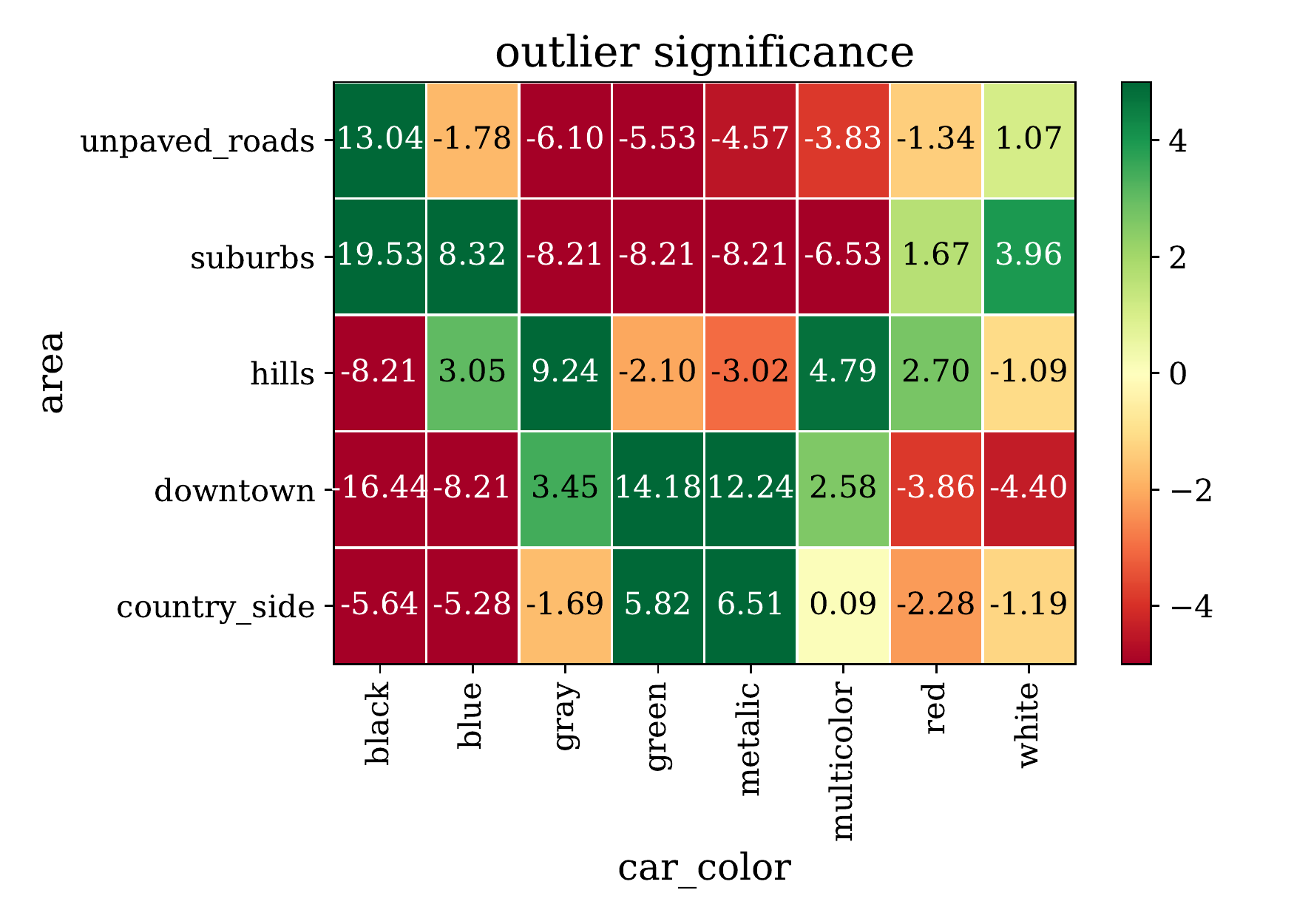}
    \caption{}
  \end{subfigure}
    \begin{subfigure}[t]{0.49\linewidth}
    \centering
    \includegraphics[width=\textwidth]{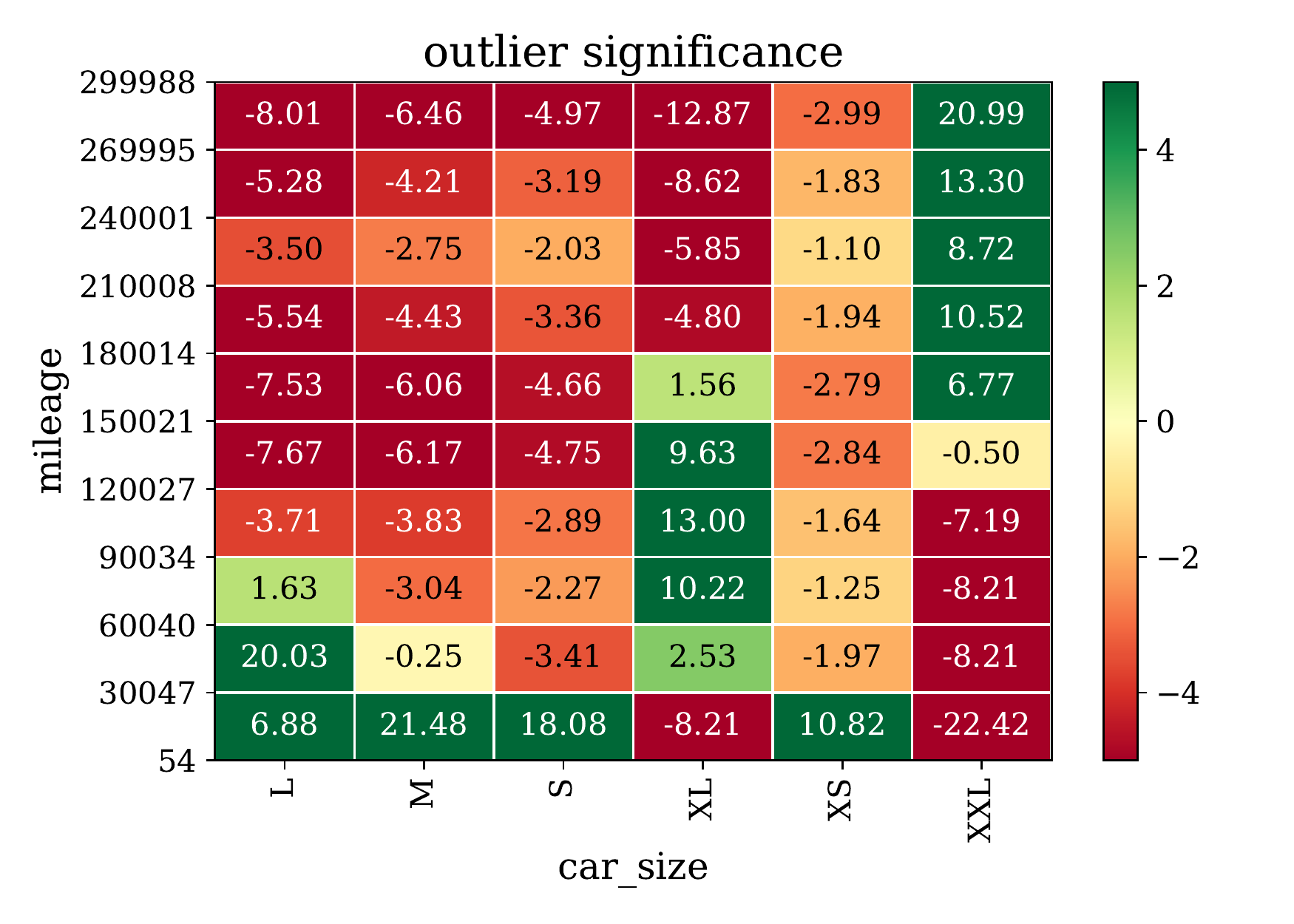}
    \caption{}
  \end{subfigure}%
    \caption{Significances of excesses or deficits of records over the expected values in a contingency table for a) the categorical variables ``car color'' and ``area''
      and b) the ordinal variable ``car size'' and the interval variable ``mileage'', measured on the synthetic data of Table~\ref{tab:data}.}
  \label{fig:contingency}
\end{figure}

Two example contingency tables are shown in Fig.~\ref{fig:contingency}, one for a combination of categorical variables, and one for
the combination of an interval and ordinal variable, both based on the synthetic car insurance data of Table~\ref{tab:data}.
Per cell each figure shows the evaluated $Z$-value.
For example, black-colored cars occur significantly more in suburbs and significantly less down-town,
and XXL-sized cars have significantly higher mileage.

In practice these turn out to be valuable plots to help interpret correlations, in particular between categorical variables.
In essence, for a data sample with a dependency, the contingency table cells with
large $|Z|$ values show the variable dependency.

\section{Three practical applications} \label{sec:applications}

Given a set of mixed-type variables and using the methods described in this work, one can:
\begin{itemize}
\item Find variable pairs that have (un)expected correlations;
\item Evaluate the statistical significance of each correlation;
\item Interpret the dependency between each pair of variables.
\end{itemize}

The methods presented in this work can be applied to many analysis problems,
and in particular they are useful for model building purposes. 
Three interesting applications using the methods presented in this paper 
are briefly discussed below.

\subsection{Modeling the frequency of insurance claims}

One interesting application is the modeling of numbers of expected insurance claims, \textit{e.g.} car damage
claims as a function of car type, type of residential area, mileage, age of driver, etc. -- a set of variables
with a mixture of types.

The aggregate loss incurred by an insurer $S$ is the total amount paid
out in claims over a fixed time period: $S = \sum_{n=1}^N s_n$, 
where $s_n$ is an individual claim amount, known as the severity, and $N$ is the total number of claims paid out in the time period.
Traditionally it is assumed that the individual claim amounts are mutually independent,
and that $N$ does not depend on the values of the claims.
The total expected severity is then expressed as a product of the expected number of claims times
the average claim amount: $E(S) = E(N)\cdot E(s)$,
where each term can be estimated separately.
When a vector of variables $\vec{x}$ is available at the individual claim level,
this information is incorporated through two independent generalized
linear models (GLMs): one for the claim frequency $N$, and the other for the severity $s$.
See Ref.~\cite{garrido_genest_schulz_2016} for more information.

Here we focus on the GLM for modeling the frequency of insurance claims.
Suppose that claims data are available for $m$ different classes of policy holders, and that class $i$ has $N_i$ claims.
Assume the claim frequency for each class is Poisson distributed, $N_i \sim P (\nu_i)$, where $\nu_i$ is the expectation for $N_i$.
Let $\vec{x}_i = (x_{i0}, ... , x_{ik})$ be the vector of variables for class $i$ at claim level, with the baseline convention that $x_{i0} \equiv 1$.
One writes:
\begin{equation} \label{eq:expnu}
\nu_i = E(N_i|\vec{x}_i) = g^{-1} (\vec{\alpha}\cdot \vec{x}_i)\,,
\end{equation}
where $\vec{\alpha} = (\alpha_{i0}, ... , \alpha_{ik})$ is a vector of regression constants\footnote{Sometimes the ratio of claims to no claims per class of policy holders is modeled instead.}.
In GLM terminology $g$ is the link function. When the frequency GLM uses a logarithmic link function, Eqn.~\ref{eq:expnu} simplifies to:
\begin{equation} \label{eq:expnu2}
\nu_i = e^{\vec{\alpha}\cdot \vec{x}_{i}}\,,
\end{equation}
yielding a simple rating structure which ensures that $\nu_i > 0$.
The logarithmic function reflects the common practice that each variable alters the baseline claim rate by a multiplicative factor.

In initial GLM risk models, no relations are typically assumed between the input variables,
and each variable category $j$ (or interval bin) is one-hot-encoded, $x_{ij} \in \{0,1\}$, and assigned one model parameter. 
The number of regression parameters per variable equals its number of categories or interval bins.
Note that it is common practice to merge low-statistics categories until they contain sufficient records.

Take the example variables of residential area and car type, each with multiple categories.
Three classes of policy holders could be: ``city, small car'', ``city, SUV'', and ``countryside, SUV'',
where the first two share the regression parameter $\alpha_{\rm city}$,
and the last two the regression parameter $\alpha_{\rm SUV}$.
The predicted, factorized number of claims for class ``city, SUV'' simply reads: $N_0\, e^{\alpha_{\rm city} + \alpha_{\rm SUV}}$,
where $N_0\equiv e^{\alpha_0}$ is the nominal number of claims shared between all classes, and $x_{\rm city} = x_{\rm SUV} = 1$.

In a refinement modeling step, to improve the factorized estimates,
cross-terms between categories of variable pairs can be added to the linear sum in the power of Eqn.~\ref{eq:expnu2}.
However, there is an exponentially large number of cross-terms to choose from. 
Practical modeling questions are: which are the most relevant terms to add? And can they be picked in an effective way that limits their number?

To help answer these, realize that the shape of Eqn.~\ref{eq:expnu2} and the assumption of variable independence  
are identical to the factorization assumption of Eqn.~\ref{eq:dep_est}.
A practical approach can then be:
\begin{enumerate}
\item Using the \phik values and their significances, select the variable pairs with the strongest correlations.
\item The most relevant model cross-terms for each variable pair $pq$, having the largest impact in the model's likelihood,
can be identified by studying the outliers in the correlation plots of Section~\ref{sec:outliers}. 
\item Cross-terms can also be included in a manner that limits the number of extra regression parameters.
For example, for a given variable pair $pq$, introduce one cross-term parameter $\beta_{pq}$ that affects only the
contingency table cells with a $Z$ value greater than a predefined value (and one for those smaller).
To model those outlier cells, use Eqn.~\ref{eq:abcderror}:
the cross term for each selected cell $i\!j$ should scale with the uncertainty on the statistically independent estimate for that cell,
$\sigma_{E_{ij}} \beta_{pq} x_{p,i} x_{q,j}$.
\end{enumerate}

\subsection{Finding unexpected answers in questionnaires}

When interpreting questionnaires one is often interested in finding all ``unexpected'' correlations
between ordinal or categorical answers given to a set of survey questions
(the definition of what constitutes an unexpected correlation is typically survey specific).
The methods presented in this paper can help to do so:
\begin{enumerate}
\item By selecting question-pairs that have an interesting (``unexpected'') $\phi_K$ correlation and significance on the one hand;
\item And selecting those with relatively high $|Z|$ values in the contingency tables of their respective answers on the other hand.
\end{enumerate}
This allows one to compile a list with all answer-pairs significantly deviating from the norm of no correlation,
of which the unexpected pairs are a subset.

\subsection{Comparison of clustering algorithms}

When looking for groups of similar data records, a typical approach is to run
multiple unsupervised clustering algorithms to cluster the data, and study the results.
In trying to understand the compatibility in clusters created by the various algorithms, 
the methods presented in this work come in useful.

For each data record, store the cluster-ID assigned by each clustering algorithm.
Using this information, one can now:
\begin{enumerate}
\item Calculate the correlation matrix between the various clustering algorithms;
\item For two specific algorithms, study where the two sets of predicted clusters overlap and deviate.
\end{enumerate}

\section{Public implementation}
\label{sec:code}

The \phik correlation analyzer code is publicly available as a Python library through the PyPi server, and from
GitHub at \url{https://github.com/KaveIO/PhiK}.
Install it with the command:

\texttt{pip install phik}

The web-page \url{https://phik.readthedocs.io} contains a description of
the source code, a tutorial on how to set up an analysis, and working examples of how to use and
run the code.

\section{Conclusion}
\label{sec:conclusion}

We have presented a new correlation coefficient, \phik, based on the $\chi^2$ contingency test, 
with Pearson-like behavior and the practical feature that it applies to all variable types alike.
Compared to Cram\'er's $\phi$, the calculation of \phik is stable against the binning per interval variable,
making it easy to interpret, and contains a noise correction against statistical fluctuations.

The asymptotic approximation breaks down for sparse and low-statistics data sets.
To evaluate the statistical significance of the hypothesis test of variable independence,
a hybrid approach is proposed where, using the $G$-test statistic, a number of Monte Carlo simulations
is used to determine the effective number of degrees of freedom and to fit an analytical, empirical description of the $\chi^2$ distribution.
We have evaluated the statistical significance of outlier frequencies with respect to the factorization assumption,
which is a helpful technique for interpreting any dependency found, \textit{e.g.} between categorical variables.

Three practical use-cases are discussed, studying the numbers of insurance claims, survey responses, and clustering compatibility,
but plenty of other applications exist.
The methods described are easy to apply through a Python analysis library that is publicly available.

\section*{Acknowledgments}
\label{sec:Acknowledgments}

We thank our colleagues of KPMG's Advanced Analytics \& Big Data team for fruitful discussions
and in particular Ralph de Wit for carefully reading the manuscript.
This work is supported by KPMG Advisory N.V.

\addcontentsline{toc}{section}{References}
\bibliography{References}{}

\end{document}